\documentclass[sigconf]{acmart}

\AtBeginDocument{%
  }

\copyrightyear{2025}
\acmYear{2025}
\setcopyright{cc}
\setcctype{by}
\acmConference[CHI '25]{CHI Conference on Human Factors in Computing Systems}{April 26-May 1, 2025}{Yokohama, Japan}
\acmBooktitle{CHI Conference on Human Factors in Computing Systems (CHI '25), April 26-May 1, 2025, Yokohama, Japan}
\acmDOI{10.1145/3706598.3713380}
\acmISBN{979-8-4007-1394-1/25/04}

\usepackage{tabu}                      
\usepackage{booktabs}                  
\usepackage{lipsum}                    
\usepackage{mwe}                       
\usepackage{soul}
\usepackage{graphicx}
\usepackage{float}
\usepackage{multirow}
\usepackage{xcolor}
\usepackage{amsmath,amssymb}
\newcommand*{\anj}{\textcolor{black}}
\definecolor{green}{RGB}{104,171, 100}
\newcommand*{\eng}{\textcolor{green}}
\definecolor{purple}{RGB}{136, 59, 139}
\newcommand*{\ab}{\textcolor{purple}}
\definecolor{blue}{RGB}{94, 154, 203}
\newcommand*{\ta}{\textcolor{blue}}
\definecolor{orange}{RGB}{254, 122, 55}
\newcommand*{\oneA}[1]{{\textcolor{orange}{#1}}}
\definecolor{brown}{RGB}{128, 53, 14}
\newcommand*{\threeA}[1]{{\textcolor{brown}{#1}}}
\definecolor{red2}{RGB}{204, 1, 2}
\newcommand*{\noT}[1]{{\textcolor{red2}{#1}}}
\definecolor{pink}{RGB}{205, 118, 168}
\newcommand*{\full}[1]{{\textcolor{pink}{#1}}}
\definecolor{olive}{RGB}{127, 126, 4}
\newcommand*{\twoA}[1]{{\textcolor{olive}{#1}}}
\definecolor{turq}{RGB}{0, 183, 176}
\newcommand*{\tO}[1]{{\textcolor{turq}{#1}}}
\definecolor{SB}{RGB}{50, 123, 168}
\definecolor{purple}{RGB}{82,37,118}
\newcommand*{\hav}[1]{{\textcolor{purple}{#1}}}
\definecolor{yellow}{RGB}{203,134,60}
\newcommand*{\lav}[1]{{\textcolor{yellow}{#1}}}

\begin{document}

\title{Lost in Translation: How Does Bilingualism Shape Reader Preferences for Annotated Charts?}

\author{Anjana Arunkumar}
\affiliation{%
  \institution{Northeastern University}
  \city{Oakland}
  \state{California}
  \country{USA}}
\email{a.arunkumar@northeastern.edu}

\author{Lace Padilla}
\affiliation{%
  \institution{Northeastern University}
  \city{Oakland}
  \state{California}
  \country{USA}}
\email{l.padilla@northeastern.edu}

\author{Chris Bryan}
\affiliation{%
  \institution{Arizona State University}
  \city{Tempe}
  \state{Arizona}
  \country{USA}}
\email{cbryan16@asu.edu}

\begin{abstract}

Visualizations are powerful tools for conveying information but often rely on accompanying text for essential context and guidance. This study investigates the impact of annotation patterns on reader preferences \anj{and comprehension accuracy} among multilingual populations, addressing a gap in visualization research. We conducted experiments with two groups fluent in English and either Tamil (\textit{n} = 557) or Arabic (\textit{n} = 539) across six visualization types, each varying in annotation volume and semantic content. \anj{Full-text annotations yielded the highest comprehension accuracy across all languages, while preferences diverged: English readers favored highly annotated charts, whereas Tamil/Arabic readers preferred full-text or minimally annotated versions. Semantic variations in annotations (L1–L4) did not significantly affect comprehension, demonstrating the robustness of text comprehension across languages. English annotations were generally preferred, with a tendency to think technically in English linked to greater aversion to non-English annotations, though this diminished among participants who regularly switched languages internally. Non-English annotations incorporating visual or external knowledge were less favored, particularly in titles. Our findings highlight cultural and educational factors influencing perceptions of visual information, underscoring the need for inclusive annotation practices for diverse linguistic audiences. All data and materials are available at: \url{https://osf.io/ckdb4/}.}
\end{abstract}



\begin{teaserfigure}
  \includegraphics[width=\textwidth]{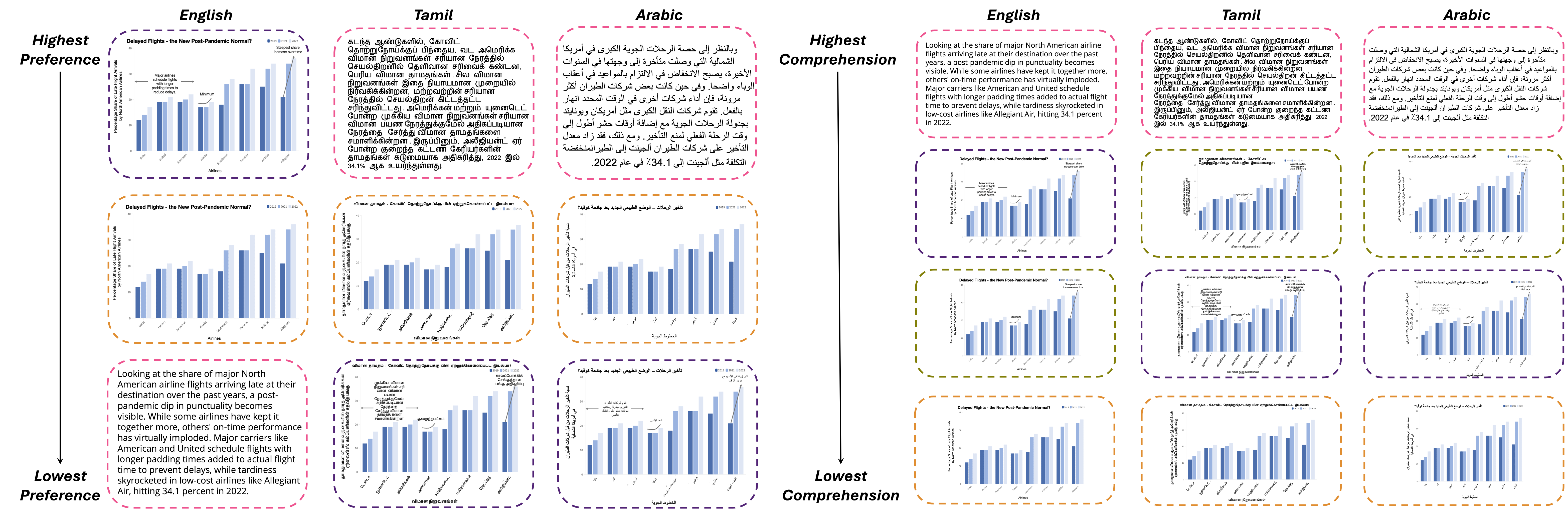}
  \centering
  \caption{
  \anj{Participants' \textbf{\textit{preferences}} and \textbf{\textit{comprehension accuracy}} for visualizations ranked across languages.} In \textbf{\textit{English}}, charts with a \textbf{\hav{high-volume}} of annotations were most preferred, though \textbf{\full{full-text}} versions had the highest comprehension accuracy. In \textbf{\textit{Tamil}} and \textit{\textbf{Arabic}}, \textbf{\full{full-text}} versions were both the most preferred and most accurate, with \textbf{\twoA{medium-volume}} annotations outperforming \textbf{\hav{high-volume}} annotations in comprehension. Charts with a \textbf{\lav{low-volume}} of annotations were moderately preferred and had low comprehension performance across all languages, though slightly ranked higher in\textbf{\textit{ Non-English}} conditions compared to \textbf{\textit{English}}.
  }
  \Description{Participants' preferences for visualizations across languages, ranked by popularity. In English, charts with a high-volume of annotation were most preferred, with preference decreasing as annotation volume declined, and full-text equivalents were least favored. In Tamil and Arabic, the pattern reversed, with full-text preferred over charts annotated at high-volumes. Low-volumes of annotation were moderately preferred across all languages, though they were higher ranked on average in Non-English compared to English.}
  \label{fig:teaser}
\end{teaserfigure}

\maketitle

\section{Introduction}

Text and symbols are crucial elements in data visualizations, providing context and enhancing interpretation~\cite{kosara2013storytelling,segel2010narrative,boy2015storytelling}. Using human-readable text in natural language is common in visualization design, primarily focusing on English annotations~\cite{bryan2020analyzing,ren2017chartaccent}. Previous studies have demonstrated that text components, such as titles, annotations, and captions, significantly influence users’ understanding and retention of visualized information~\cite{borkin2013makes,10294209,borkin2015beyond}. As a result, researchers have established best practices for optimizing the volume, content, and placement of text annotations to enhance communication in visualizations~\cite{brath2020literal,stokes2022striking,stokes2023role}.

However, much of this research is grounded in the context of English-speaking WEIRD (Western, Educated, Industrialized, Rich, and Democratic) societies~\cite{linxen2021weird}, overlooking the diverse linguistic and cultural contexts in which visualizations are utilized globally~\cite{rakotondravony2023beyond,zhao2016cross,boyd2021can}. This focus on English and monolingual settings can marginalize non-English-speaking communities and limit our understanding of how different linguistic backgrounds affect interactions with visual data~\cite{fernandez2021underrepresented,eddens2017interactive}.

There are several situations in which multilingualism introduces additional complexity to data visualization that warrants dedicated exploration, including differences in language use across contexts, variations in cognitive processing between languages, and affective responses to language. For example, in many contexts, distinct languages are used for different purposes, such as colloquial communication and formal instruction~\cite{guthrie2015formalistic,pfeiffer2023language}. Additionally, research in cognitive psychology and neuroscience indicates that using a second language can influence cognitive processing, often prompting more analytical and less emotional responses due to increased cognitive load and the engagement of the prefrontal cortex~\cite{keysar2012foreign,grundy2017neural,marian2017bilingual}. This cognitive shift may alter how multilingual individuals interpret visual information and respond to annotations, potentially leading to different preferences compared to monolingual individuals. Language context may also shape trust and engagement with visualizations; multilingual individuals might perceive and trust information differently when it is presented in a non-native language due to factors such as perceived credibility, familiarity, or cognitive effort required~\cite{duarte2020experiences,henderson2011does}.
Given the underexplored complexities of working with multilingual audiences, it is possible that current annotation recommendations may not generalize across languages. Thus, it is critical to explore how multilingual contexts intersect with visualization design to ensure inclusivity and accessibility.

Our study explores the preferences of bilingual individuals for annotated data visualizations, aiming to understand how prior work on visualization annotations generalizes to diverse contexts.  While this paper discusses multilingual contexts broadly, our study specifically focuses on individuals who are fluent in English and either Tamil (\textit{n} = 557) or Arabic (\textit{n} = 539). This focus on bilingualism allows us to investigate how fluency in two languages, both with distinct linguistic structures and cultural contexts, affects interpretation and preferences in data visualization. 

\anj{We selected Tamil and Arabic for this study due to their linguistic diversity, global significance, and their substantial bilingual populations that speak English as a second language. Tamil, a Dravidian language spoken by over 75 million people worldwide~\cite{zhang2021multilingual}, and Arabic, a Semitic language spoken by over 300 million people~\cite{guellil2021arabic}, represent two major linguistic families that are vastly different from English in terms of grammar structure, script, and cultural context~\cite{sainulabdeen2022contrastive}. Moreover, both Tamil and Arabic are official languages in multilingual societies (India, Sri Lanka, and various countries in the Middle East and North Africa), where bilingualism or multilingualism is common~\cite{sarveswaran2024morphology,najjar2020teaching}. Their prevalence and bilingual populations makes them reflective of many bilingual audiences who regularly switch between languages in educational, professional, and everyday contexts~\cite{nj2020investigation,pubadi2020focus,adder2020english}.}

Using a ranking task, we elicited participants' \textbf{\textit{preferences}} for varying annotation patterns across languages and additionally evaluated their \textbf{\textit{comprehension}} of the visualization by having them select correct conclusions. We tested both preferences and comprehension using six types of visualizations-- bar charts, pie charts, line graphs, maps, scatter plots, and heatmaps-- with varying annotation volume and semantic content, in both their native language and in English. Participants also provided free-response feedback on their \textbf{\textit{likes and dislikes}} about the stimuli presented.

Specifically, we contribute:
\begin{itemize}
    \item Design exploration for creating bilingual data visualizations, including the adoption and application of annotation placement techniques, translation processes, and visual consistency checks across English, Tamil, and Arabic (Sec.~\ref{sec:4})
    \item Quantitative findings on preferences for varying annotation volumes in bilingual individuals, and insights into how linguistic immersion influences individual preferences (Sec.~\ref{sec:6})
    \item Qualitative results of why readers prefer certain annotation patterns in different languages and how they utilize the information presented (Sec.~\ref{sec:7})
    \item Recommendations for designing inclusive and culturally sensitive visualizations, focusing on optimizing text annotation practices in multilingual contexts (Sec.~\ref{sec:8},~\ref{sec:9})
\end{itemize}

\anj{As a preview, our findings reveal a general preference for higher volumes of English annotations across all chart types, contrasted with a preference for fewer annotations when content is presented in native languages, as shown in Figure~\ref{fig:teaser}. Specifically, participants preferred English annotations when interpreting complex and detailed information, while favoring concise annotations in their native language. A tendency to think technically in English was linked to a stronger preference for English over non-English annotations, though this effect diminished among individuals who frequently switch between languages in their internal monologue. Beyond preferences, our study also investigated how bilingual individuals interpret annotations in their native and second languages, focusing on the roles of annotation volume and semantic content in shaping comprehension accuracy. While full-text annotations consistently resulted in the highest comprehension accuracy across all languages, semantic variations (L1–L4 levels) did not significantly impact comprehension. These results suggest that preferences, rather than comprehension, may drive the choice of annotation styles for multilingual audiences. Together, our findings highlight the critical role of both semantic content and linguistic immersion in shaping annotation preferences and comprehension, emphasizing the need for adaptable and culturally sensitive design practices in multilingual contexts.}

\section{Background}

Our work builds on research in text integration, multilingual cognition, and linguistic and cultural factors in visualizations. We examine how annotations in different languages affect reader preferences and comprehension. To support diverse readers, we briefly review relevant theories in multimedia learning, cognitive load, and the impact of bilingualism on visual interpretation.

\subsection{Integrating Text and Charts}

A growing body of research shows that integrating text with visualizations significantly influences readers' conclusions and information recall~\cite{kim2021towards,kong2018frames,kong2019trust,stokes2022striking}. For example, Borkin et al.\cite{borkin2013makes,borkin2015beyond} found that participants focus more on textual elements like titles and labels, while Kong et al.\cite{kong2018frames} demonstrated that a chart's title could shape readers' memory of its content. Additionally, captions~\cite{kim2021towards} and annotations~\cite{stokes2022striking} direct interpretations by providing context. Arunkumar et al.~\cite{10294209} found that minimal-text visualizations are rated higher for affective engagement, while text-heavy visuals are perceived as more informative and cognitively stimulating. Overall, these studies highlight text’s crucial role in shaping the effectiveness of visualizations.

One theory that could explain part of the impact of text in visualization is Mayer et al.'s work on the \textit{multimedia effect} posits that people learn more effectively when words and images are combined, rather than used alone~\cite{mayer2020multimedia}. Supporting research shows that placing text near explanatory images reduces cognitive load and improves information processing~\cite{holsanova2005tracing,holsanova2009reading,zhao2014eye}. While the multimedia effect suggests positive outcomes from dual encoding, the relationship between text and visuals may be more complex. For instance, studies have found that annotations enhance preferences and engagement with charts~\cite{stokes2022striking}, yet others, like Hearst and Tory~\cite{hearst2019would}, indicate that some users prefer to omit visuals, highlighting a more nuanced interaction dependent on user preferences and tasks.

\begin{table}[!ht]
    \centering
    {
    \scriptsize
    \begin{tabular}{p{0.05\textwidth}p{0.2\textwidth}p{0.15\textwidth}}
        \toprule
        \textbf{Semantic Level}   &   \textbf{Description} & \textbf{Example (From Fig.~\ref{fig:seta})}  \\ 
        \midrule
                \textbf{L1}    &    Consists of elemental or encoded aspects of the chart, such as the overall topic or a description of the content of an axis. &   \textit{``2010 Teenage Birth Rates in America (Per 1000 Women)"}
                \\ 
                \textbf{L2}    &    Consists of statistical or relational components, such as a comparison between two points or identification of extrema. &  \textit{``Minimum"}         
                \\ 
                \textbf{L3}    &    Describes perceptual or cognitive aspects, such as an overall pattern or changes in trend. &   \textit{``Steepest overall fall, followed by a more gradual decrease"} \\   
                \textbf{L4 }   &    Provides external context to the chart, such as past events which affect the topic depicted. &   \textit{``Cultural and religious beliefs, access to medical services and lack of financial resources continue to hamper progress"}    
                \\ 
                \bottomrule                                         
    \end{tabular}%
}
\caption{Conceptual model of the semantic content of chart annotations by Lundgard et al.~\cite{lundgard2021accessible} with corresponding examples in Fig.~\ref{fig:seta}.}
    \label{tab:1}
    \vspace{-6mm}
\end{table}

We are not the first to explore the nuanced relationship between text and visuals. Lundgard et al. \cite{lundgard2021accessible} developed a conceptual model with four levels of semantic content (see Table~\ref{tab:1}) to design alt-text descriptions for charts used by screen readers. They found that while sighted readers preferred high-level explanations and domain-specific context, blind and low-vision users favored mid-level descriptions of statistical features or perceptual trends over abstract explanations. These findings challenge the idea that text simply serves as dual encoding, emphasizing the need to better understand how text and visuals interact across contexts and user groups.

\subsection{Linguistic and Cultural Considerations in Chart Design}
\label{sec:rakoto}

\anj{Building on the critical role of text in visualizations, linguistic and cultural considerations are equally crucial in chart design, influencing how information is communicated and interpreted across diverse audiences~\cite{rakotondravony2023beyond,9904569}. Studies have highlighted the importance of linguistic factors such as language proficiency~\cite{golubeva2021impact}, readability~\cite{midway2020principles}, and clarity~\cite{burns2020evaluate} in communicating information. Assessing reasoning with charts also requires accounting for individual differences in visual literacy, influenced by language and culture~\cite{messaris2018visual,treffers2020explaining}. For example, Correll~\cite{10.1145/3290605.3300418} highlights the potential alienation viewers may feel from data represented in standard visualizations. 
Similarly, Peck et al.~\cite{peck2019data} highlight that one-size-fits-all approaches in visualization research often overlook certain demographic groups, emphasizing the role of individuals' experiences in shaping their attitudes toward visualizations. These studies stress the need to rethink research practices to ensure inclusivity and consider diverse audiences. Further research is needed to explore the interaction between text and visualization in underrepresented languages, especially as natural language interfaces for data visualization gain traction~\cite{10541799,dibia2023lida}.}

\anj{In the context of chart design considerations, Rakotondravony et al.~\cite{rakotondravony2023beyond} shows how the verbalization of quantitative probability through visualizations can vary across languages and how non-English languages interplay with data visualization reasoning in Madagascar. Alebri et al.\cite{alebri2024design} extend this by identifying a mix of Right-to-Left (RTL) and Left-to-Right (LTR) design approaches used for visualizations in Arabic media. Notably, they found that while RTL orientations are more commonly applied to categorical data, there is significant variability in design practices, with other visualization types inconsistently utilizing both RTL and LTR orientations within the same article. This inconsistency reflects the lack of standardized RTL design guidelines, especially for charts translated or adapted from English sources. }

\anj{While our study incorporates Arabic-language annotations, we intentionally focus on text preferences and comprehension without altering chart orientations. This choice acknowledges the complexity of right-to-left adaptation and aligns with Alebri et al.'s findings that such modifications require additional considerations beyond our current scope. Instead, we aim to investigate text preferences independently of directional effects, especially for charts like maps and pie charts, where axes are absent.}

\subsection{Bilingualism and Cognition in Visualization Interpretation}

Understanding how bilingualism affects cognitive processing is crucial for designing visualizations that cater to multilingual audiences. Research in cognitive psychology and neuroscience shows that using a second language often requires more cognitive resources and engages different brain regions, particularly those involved in higher-order functions like planning and problem-solving~\cite{keysar2012foreign,grundy2017neural}. This increased cognitive load can lead to more analytical and less intuitive decision-making processes~\cite{marian2017bilingual}. For instance, individuals presented with moral dilemmas in a second language tend to make more utilitarian, rational choices, suggesting that second language processing can dampen emotional resonance and lead to more de-

\begin{table*}[!ht]
\centering
\scriptsize
\begin{tabular}{@{}lllll@{}}
\toprule
\textbf{Task} & \textbf{Study Phase} & \textbf{Stimuli Sets Used} & \textbf{RQ Addressed} & \textbf{Effect Tested}\\ 
\midrule
Reliance Ratings & \textbf{1} &  Set A & \textit{RQ2} & Quantitative analysis of comprehension based on annotation \textit{density}\\
 & \textbf{3} &  Set B & \textit{RQ2} & Quantitative analysis of comprehension based on annotation \textit{content}\\
Conclusion Selection Accuracy & \textbf{1} &  Set A & \textit{RQ2} & Quantitative analysis of comprehension based on annotation \textit{density}\\
 & \textbf{3} &  Set B & \textit{RQ2} & Quantitative analysis of comprehension based on annotation \textit{content}\\
Free Response Feedback & \textbf{1} &  Set A  & \textit{RQ1} & Qualitative analysis of preferences based on annotation \textit{density}\\
 & \textbf{3} &  Set B & \textit{RQ1} & Qualitative analysis of preferences based on annotation \textit{content}\\
Rankings & \textbf{2} &  Set A & \textit{RQ1} & Quantitative analysis of preferences based on annotation \textit{density}\\
 & \textbf{4} &  Set B & \textit{RQ1} & Quantitative analysis of preferences based on annotation \textit{content}\\
Linguistic Immersion Ratings & \textbf{5} & N/A & \textit{RQ3} & Moderating effect of \textit{linguistic individual differences} on preferences and comprehension\\ 
\bottomrule
\end{tabular}%
\caption{\anj{Table summarizing study phases, their constituent tasks (detailed in Sec.~\ref{sec:method}), and the associated RQs.}}
\label{tab:rqhmap}
\end{table*}

\noindent tached problem-solving~\cite{costa2014your, hayakawa2017thinking}.

In the context of data visualization, this cognitive shift has important implications. The increased cognitive effort required for processing information in a second language can influence how multilingual individuals interpret visual information and respond to text annotations. For example, research has shown that bilinguals often exhibit different reading patterns when processing information in their non-native language, such as increased fixation durations and more frequent backtracking when reading text~\cite{cop2015eye, marks2022integrated, jasinska2014development}. These differences in reading behavior indicate that bilingual individuals might require more time to integrate textual and visual information, potentially affecting their preferences for the quantity and type of annotations in data visualizations~\cite{rayner200935th, van2008language}.

Moreover, the mental effort associated with second language use can influence how effectively individuals process and retain information presented in visualizations. Enhanced cognitive load can reduce working memory capacity, limiting the ability to simultaneously process visual data and accompanying text~\cite{baddeley2003working, conway2003working}. This is particularly relevant in complex visualizations where the integration of multiple information sources is necessary. As a result, bilinguals may prefer simpler, more straightforward visualizations with fewer text annotations when using a second language, as these are less cognitively demanding~\cite{sweller2011cognitive, vohs2016handbook}.

Furthermore, bilingualism itself introduces unique cognitive dynamics. Bilinguals often switch between languages depending on the context, which can enhance cognitive flexibility and the ability to manage multiple streams of information simultaneously~\cite{bialystok2009bilingualism, hernandez2013bilingual}. This cognitive flexibility might make bilinguals more adept at navigating visualizations that integrate both textual and visual elements, although the cognitive cost of language switching should not be underestimated~\cite{declerck2013bilingual, kramer2015effects, linck2012inhibitory}. Such switching could potentially alter the depth of information processing, with bilinguals perhaps relying more on visual cues than on textual information when presented in their non-dominant language~\cite{timmermeister2020no}.

Given these cognitive considerations, our study focuses on bilingual individuals fluent in English and either Tamil or Arabic to explore how language fluency affects interpretation and preferences in data visualization. By examining both linguistic structures and cultural contexts, we aim to understand how bilinguals process visual and textual information and how this affects their engagement with different types of visualizations (see Section~\ref{sec:5}).
\vspace{-3mm}

\section{Methodology}
\label{sec:5}

\anj{To begin to examine the relationship between text and visual representations of data for bilingual audiences, we designed a study with five phases, aligning each phase with specific research questions (RQs) and hypotheses (Hs), as shown in Table~\ref{tab:rqhmap} (see Sec.~\ref{sec:rqs} for detailed descriptions of RQs). This design ensures that each RQ is systematically addressed, covering preferences, comprehension, and individual differences. Below, we briefly describe these RQs and how they connect to our methodology:}

\anj{\noindent \textbf{RQ1 (Preferences)}: Bilingual readers’ annotation preferences are evaluated through \textbf{free-response} tasks (Phase 1) and \textbf{ranking} tasks (Phases 2 and 4) on charts annotated in native (Arabic/Tamil) vs. second-language (English). These phases test \textbf{H1}, predicting that participants will show a stronger preference for native language annotations, with this preference increasing as annotation volume grows.}

\anj{\noindent \textbf{RQ2 (Comprehension)}: Bilingual readers' ability to interpret annotations is assessed through \textbf{comprehension} tasks (Phase 3), where participants select correct conclusions about charts from a list. Additionally, self-reported \textbf{reliance on text} versus visual elements (Phase 1) is analyzed to examine its influence on comprehension. These phases test \textbf{H2}, which predicts higher comprehension accuracy for annotations in the native language, particularly for complex charts.}

\anj{\noindent \textbf{RQ3 (Individual Differences)}: Individual \textbf{linguistic differences} are collected through a demographic survey (Phase 5), capturing linguistic immersion, educational background, and fluency levels. These data are analyzed to test \textbf{H3}, which explores how linguistic immersion impacts annotation preferences and comprehension accuracy.}

\begin{figure}[H]
\centering
    \includegraphics[width=0.5\textwidth]{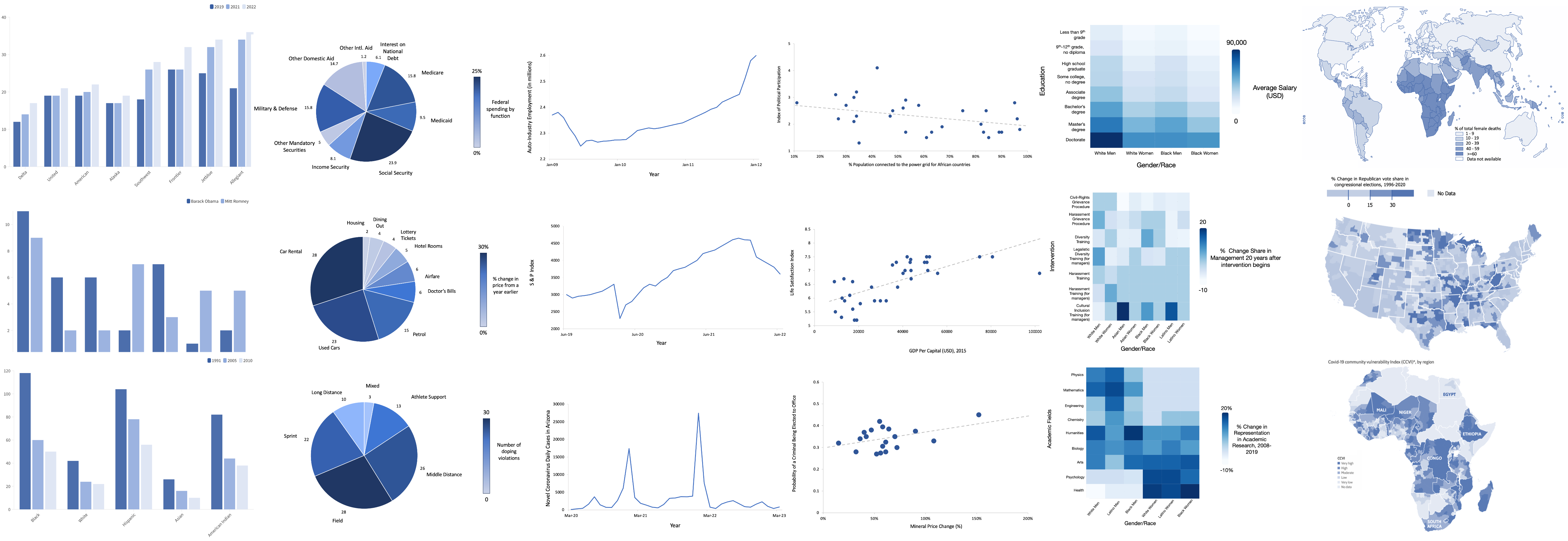}
    \caption{\anj{18 charts generated for study phases 1--4 (unannotated), spanning 6 chart types x 3 data shapes. In our main study, participants are shown one data shape per chart type-- i.e., one chart per column from this figure-- to complete study tasks. This was done to reduce the potential biasing effect of data shape.}}
    \Description{18 charts generated for this study (unannotated), spanning six chart types: bar charts, pie charts, line charts, scatterplots, heatmaps, and maps.}
    \label{fig:blankstimuli}  
\end{figure}

\begin{figure*}[!ht]
    \centering
    \includegraphics[width=\textwidth]{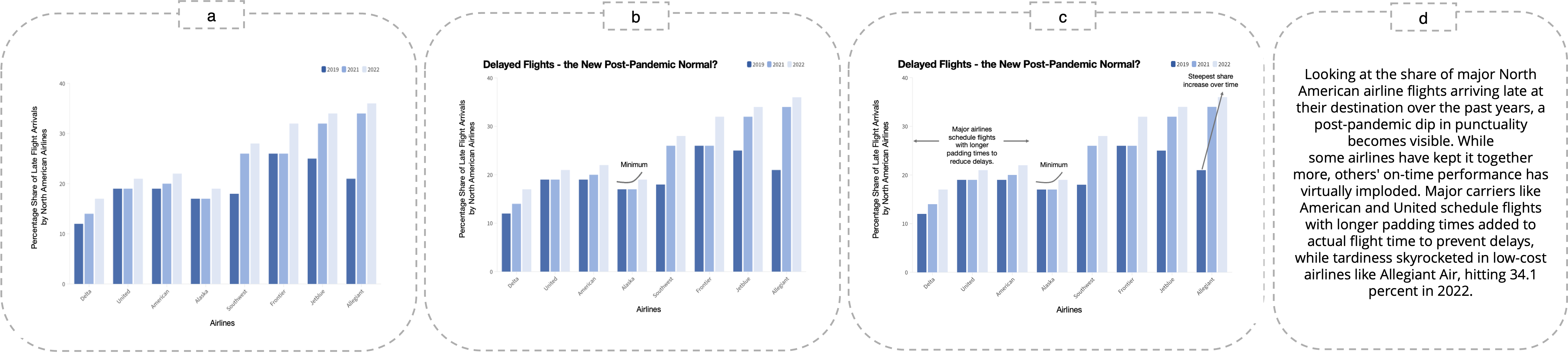}
    \caption{\anj{Example bar chart stimuli for Set A, annotated in English, a total of \textbf{4 variants}. We aimed to capture \textit{preferences} \textbf{(RQ1)} and \textit{comprehension} \textbf{(RQ2)} for the extremes between visual and textual presentation of information. (a) Chart presented with no text (beyond axes and ticks). (b) Chart with a title and a single annotation (Title + 1A). (c) Chart which displays a narrative or story around the data, annotated through text (Title: L1 + 3 annotations: L2--L4; Title + 3A). (d) A text-only version of the data, with the same story as displayed in (c).}}
    \Description{Example of 4 bar chart stimuli for Set A, annotated in English. Information displays with varying amounts of text annotation across charts.}
    \label{fig:seta}  
\end{figure*}

\begin{figure*}[!ht]
\centering
    \includegraphics[width=\textwidth]{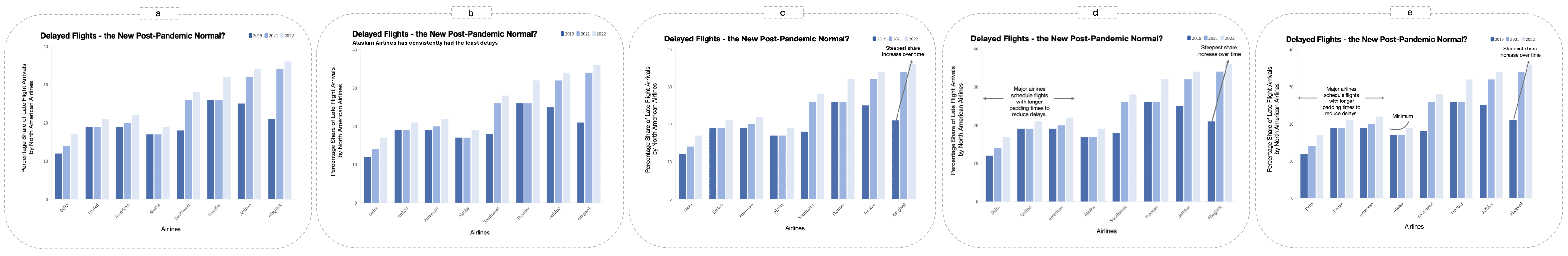}
    \caption{\anj{Example bar chart stimuli for Set B (fine-grained comparisons), annotated in English, a total of \textbf{11 variants}. We aimed to capture \textit{preferences} \textbf{(RQ1)} and \textit{comprehension} \textbf{(RQ2)} for different levels of semantic content in chart annotations. We accordingly construct variants such that for different annotation volumes, different combinations of L2, L3, and L4 annotations may be present. (a), (b) \textbf{Title-Only} charts: the main title represents L1 and subtitles are included to incorporate L2--L4 information (total: 4 variants). (c) \textbf{Title+1A} charts: title + a single embedded annotation from L2--L4 (total: 3 variants). (d) \textbf{Title+2A} charts: title + two embedded annotations from L2--L4 (total: 3 variants). (e)  \textbf{Title+3A} chart: title + three embedded annotations from L2--L4 (total: 1 variant).} 
}
\Description{Example of 5 bar chart stimuli for Set B, annotated in English. Information displays with varying amounts and semantic content of text annotation across charts.}
    \label{fig:setb} 
\end{figure*}

\anj{Our stimuli include six chart types (bar charts, pie charts, scatterplots, line charts, maps, and heatmaps), chosen to represent a range of visual encodings and complexity levels. For instance, scatterplots and heatmaps represent bivariate data, while the remaining charts are univariate. Each chart type is created in three variations (as shown in Fig.~\ref{fig:blankstimuli}) by manipulating the underlying data (i.e., changing the \textit{data shape}). Participants view one variation per chart type at random in the study, ensuring that annotation content and volume, rather than data patterns, drive responses. This randomization helps neutralize potential biases introduced by visual differences in data patterns. The charts are annotated with varying text densities and semantic structures in three languages: English, Tamil, and Arabic. We discuss the stimuli creation process in detail in Sec.~\ref{sec:4}.}


\subsection{Stimuli}
\label{sec:4}

\anj{
To test the research questions, we devised two sets of stimuli. We adapted a similar approach to Kim et al.~\cite{kim2021towards} and Stokes et al.~\cite{stokes2022striking}, to design our base charts and place annotations to create the variants in both sets. Essentially, our stimuli design extends Stokes et al.'s work to a bilingual context, testing how semantic levels interact with language (native vs. second) to influence \textit{preferences} \textbf{(RQ1)} and \textit{comprehension} \textbf{(RQ2)}}.

\anj{
\textbf{Set A} tested the premise that bilingual readers \textit{preferred} \textbf{(RQ1)} native language annotations, and that both \textit{preference} \textbf{(RQ1)} and \textit{comprehension} \textbf{(RQ2)} strengthened for \textit{higher annotation volumes}. This set included four variants of text annotation volume, ranging from a chart with no text (except axes; Fig.~\ref{fig:seta} A) to a text paragraph with no chart (Fig.~\ref{fig:seta} D).
}

\anj{
\textbf{Set B} examined finer differences in how \textit{annotation content}, i.e., the semantic level of annotations~\cite{lundgard2021accessible,stokes2022striking} impacted chart \textit{preference} \textbf{(RQ1)} and \textit{comprehension} \textbf{(RQ2)}. This set comprised 11 variants of annotation content, ranging from a chart with only a title to a chart with a title and three annotations (see Fig.~\ref{fig:setb}). 
}

\begin{figure*}[!ht]
\centering
    \includegraphics[width=0.8\textwidth]{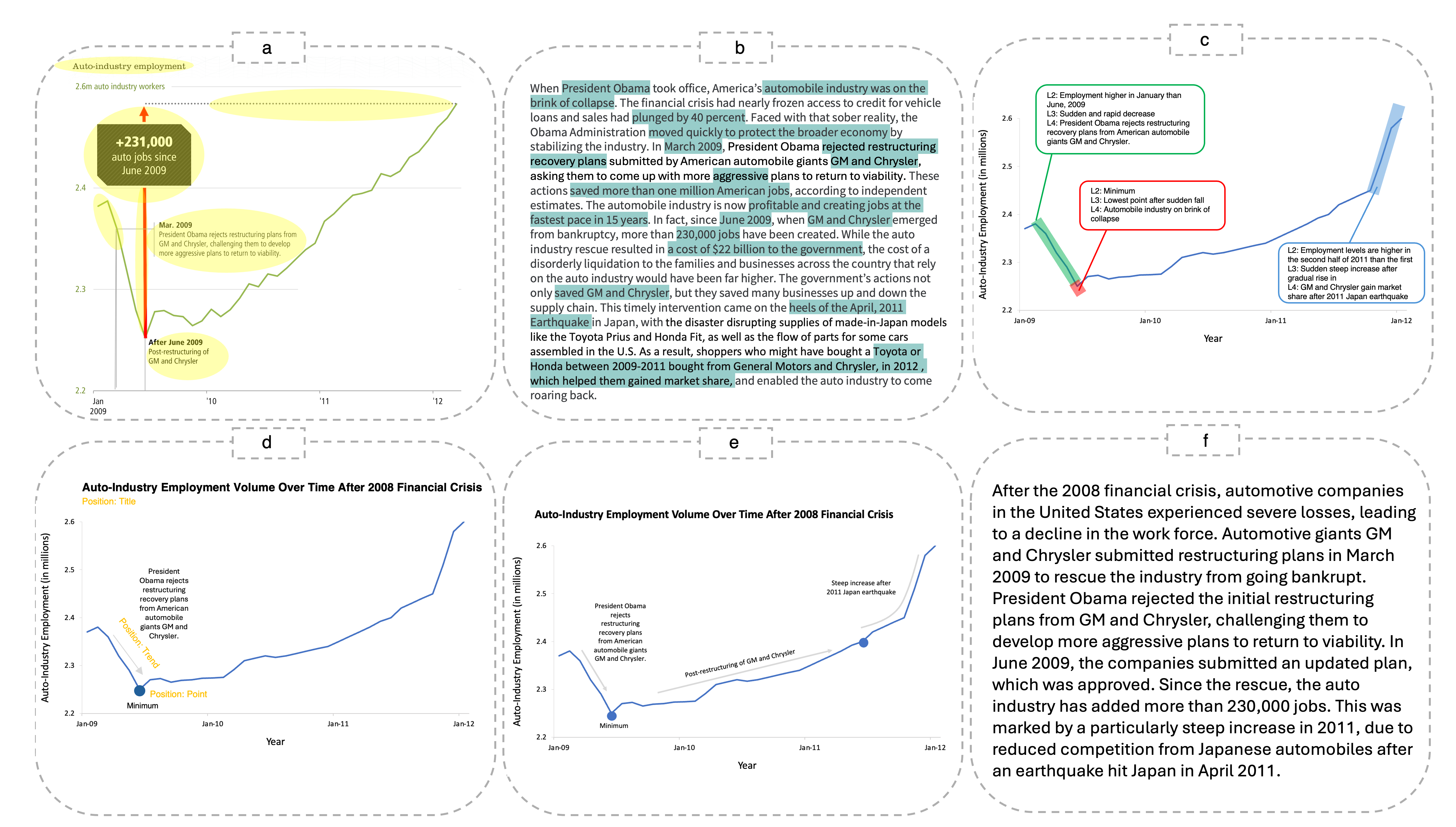}
    \caption{Example of the stimulus creation process, based on the identification, ranking, and synthesis of chart/text emphasis features. The initial chart from the article is shown in (a), with the most prominent visual emphasis features highlighted in yellow. (b) represents the corresponding article text, which has textual emphasis features highlighted in teal. In (c), the blank chart created using \texttt{d3} is shown, with potential annotations for highlighted prominent regions. Red indicates the most prominent region, green the second, and blue the third. The types of annotation positions are outlined in (d). In step (e), an expert designer adjusted the fine details to produce a chart with a realistic layout, corresponding to (f) a summary text paragraph synthesized by the annotators.}
    \Description{Example of the stimulus creation process, based on the identification, ranking, and synthesis of chart/text emphasis features. The initial chart from the article is shown, with the most prominent visual emphasis features highlighted in yellow. The corresponding article text has textual emphasis features highlighted in teal. A blank chart is created using d3, with potential annotations for highlighted prominent regions. Red indicates the most prominent region, green the second, and blue the third. An expert designer adjusted the fine details to produce a chart with a realistic layout, corresponding to a summary text paragraph synthesized by the annotators.}
    \label{fig:caa-process} 
\end{figure*}

\subsubsection{Design Process}
\hfill\\

\noindent \anj{\textbf{Underlying Data}: First, we selected real-world chart-article pairs from prominent news media (Economist, New York Times, The BBC), covering a range of topics, including economics, healthcare, education, and environment. For each chart type, we selected three pairs with consistent types across articles and charts. This offered a balance between everyday relevance (e.g., healthcare trends) and topics requiring specialized knowledge (e.g., economic indices), controlling for bias from overly familiar or obscure topics. Crucially, none of the topics were highly polarizing or controversial, to avoid the confounding effects of personal or political biases. This aligned with prior research showing that user interaction with visualizations can be influenced by topic familiarity and engagement~\cite{hullman2015content, kim2021towards}.}

\noindent\anj{\textbf{Base Chart Construction}: We then used \texttt{d3.js}~\cite{bostock2011d3} to recreate the basic, text-free charts. We included six types of charts: bar, pie, line, map, scatterplot, and heatmap. We considered these because they are among the most common basic charts, and their underlying data contain either temporal- or frequency-based features that can be easily annotated. Each chart type comprised a subset of the underlying data from three unique articles, resulting in three unique data shapes, each of which contained at most two trends. This ensured relatively realistic global shapes with sufficient variation in the stimuli while maintaining enough blank space on the chart in which to situate textual and visual annotations. The full set of generated charts can be seen in Fig.~\ref{fig:blankstimuli}.}

\noindent\anj{\textbf{Chart Annotations}: We then performed the following annotation process (as illustrated in Fig.~\ref{fig:caa-process})\footnote{See Supplemental Material for annotated versions of the real-world charts and article text.} by recruiting five annotators who are researchers in data visualization:}

\textbf{(A) Chart Emphasis:} Annotators independently labeled visually prominent features for each chart image in the dataset. Later, these labels were merged to reach a consensus. To prevent bias, annotators refrained from reading any article text during this process~\cite{kim2021towards}. 

\textbf{(B) Text Emphasis: } Annotators identified paragraphs in each article that contained information relevant to the chart content, complemented prominent or non-prominent chart features, or reiterated annotations already present on the charts. Subsequently, the lists of identified text features were merged to reach a consensus.

\anj{\textbf{(C) Annotation Creation:} Annotators ranked the identified chart and text emphasis features in order of importance. The top three ranked prominent chart emphasis features, corresponding to visual data marks on the respective charts, were considered to be the three chart regions for annotation placement. Text emphasis features from the article were synthesized into single-sentence annotations suitable for placement on the chart, either as a title or a label. Subsequently, the created annotations were semantically categorized based on Lundgard et al.'s~\cite{lundgard2021accessible} conceptual model, as summarized in Table~\ref{tab:1}, with annotator consensus. The top-ranked annotations (one per semantic category) were selected to be applied to the chart annotation regions, as shown in Figure~\ref{fig:caa-process} (C). Following Stokes et al.~\cite{stokes2022striking}, annotations were left-justified and non-overlapping with chart marks, which were presented in blue (\textcolor{SB}{$\blacksquare$}) to blend with the survey instrument; stimuli met at least AA guidelines under WCAG 2.0 testing. Additionally, for each chart in the stimuli set, we included an all-text variant, following Stokes and Hearst~\cite{stokes2021give}; annotators synthesized these as single-paragraph descriptions by combining chart annotations with additional text content from the article. }

\begin{figure*}[!ht]
\centering
    \includegraphics[width=0.8\textwidth]{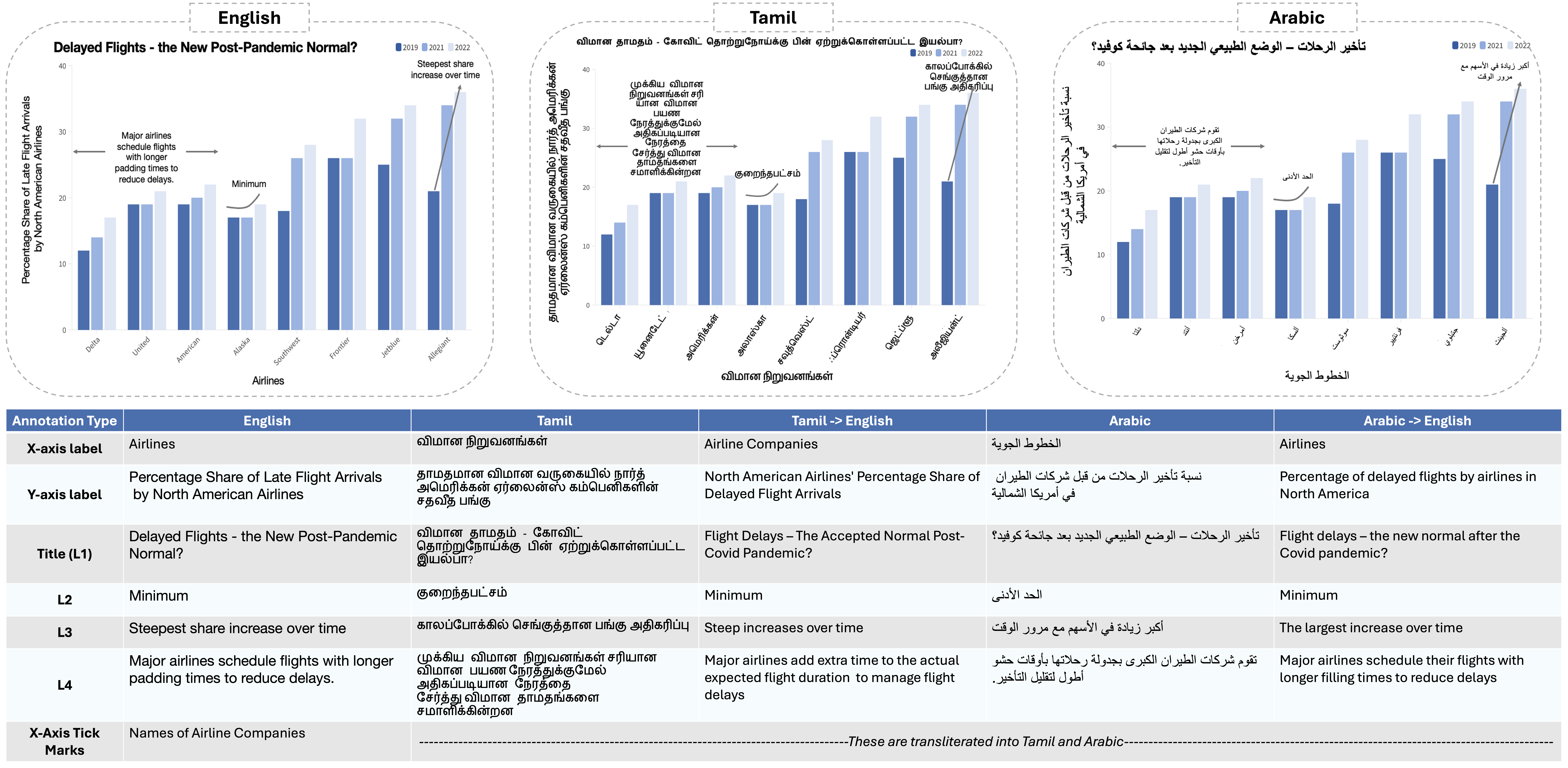}
    \caption{Example of a bar chart translated into Tamil and Arabic. The table displays the back-translations of Tamil and Arabic into English, for all chart annotations. Note that the x-axis tick marks are transliterated phonetically as they represent the names of airline companies.}
    \Description{Example of a bar chart translated into Tamil and Arabic, with a table below showing the back-translations of Tamil and Arabic into English for each chart annotation element.}
    \label{fig:caa-translate} 
\end{figure*}

\anj{The resultant chart variants produced comprise stimuli sets A, B for the English language, which are used to test chart \textit{preferences} \textbf{(RQ1)} and \textit{comprehension} \textbf{(RQ2} for English annotations in our study.}

\anj{\textbf{Translating Charts to Tamil and Arabic}: 
A group of six international student volunteers from \textit{Anonymized University} then translated all the chart annotations and full-text variants from Set A and Set B into Tamil and Arabic. Working independently, three annotators per language compared and merged translations by consensus, following cross-linguistic guidelines, including back-translation and semantic validation~\cite{beaton2000guidelines}. In the back-translation process~\cite{brislin1970back}, after the text was translated from English into Tamil or Arabic, it was retranslated back into English by a different translator who had not seen the original text. The back-translated version was then compared with the original English text to identify any differences in meaning, tone, or cultural nuances.
Semantic validation involved a thorough review of the translated texts to ensure that the intended meaning and contextual relevance were preserved across languages~\cite{guillemin1993cross,tariq2016semantic}.}

\anj{Annotators compared translations for linguistic accuracy and cultural appropriateness, ensuring clarity, conciseness, and alignment with the original concepts. Accurate and meaningful translation of the content was prioritized over directly controlling for text volume differences as linguistic structures naturally result in varying text lengths across languages; on average, annotation length increased by 36\% for Tamil and 14\% for Arabic when translated from English. This approach reflects real-world design practices, as described by Alebri et al.~\cite{alebri2024design}, where replicating the semantic and contextual fidelity of the original content takes precedence over achieving uniformity in text volume. Our translation process involved multiple review iterations to ensure cultural relevance and linguistic fidelity, with annotators reaching a consensus before finalizing each version. Fig.\ref{fig:caa-translate} provides an example of a translated chart in all three languages, while Figs.\ref{fig:seta} and \ref{fig:setb} show the final annotated chart versions used as study stimuli (in English). This rigorous approach allowed us to maintain the integrity of the content while addressing the challenges of multilingual design.}

\anj{We also note that while our stimuli include Arabic--language annotations, we did not adapt the charts' orientations to RTL (right-to-left) layouts. This decision aligns with prior findings by Alebri et al.~\cite{alebri2024design} (as mentioned in Sec.~\ref{sec:rakoto}) that designers in Arabic news outlets predominantly employ RTL orientations for categorical data, but mix RTL and LTR (left-to-right) approaches for other data types, often inconsistently within the same chart or article. In our study, we utilized visualizations translated from English media sources, which inherently follow LTR design conventions. Given the lack of standardized practices for RTL visualizations, as identified by Alebri et al., we chose to maintain the original LTR orientation in our stimuli. This decision was made to preserve the integrity of the original design and to avoid introducing additional variables that could confound our results. While this approach may not fully align with the reading habits of RTL language readers, it allows for a controlled examination of how language translation affects preferences and comprehension without the added complexity of varying design orientations. Additionally, by studying both types of stimuli-- those with axes (line, bar, scatterplot, heatmap) and those without axes (map, pie)--we ensure that our findings are generalizable to a broad range of visualization contexts.}

\anj{The resultant chart variants produced comprise stimuli sets A, B for the Tamil/Arabic languages, which are used to test chart \textit{preferences} \textbf{(RQ1)} and \textit{comprehension} \textbf{(RQ2} for native language annotations in our study.}

\vspace{-4mm}
\subsection{Research Questions}
\label{sec:rqs}

To begin to examine the relationship between text and visual representations of data for bilingual audiences, we explored three primary research questions outlined below, focusing on user \textbf{\textit{preferences}} (RQ1),\textbf{\textit{ individual differences}} (RQ2), and \textbf{\textit{comprehension}} (RQ3). Our analysis, detailed in Section~\ref{sec:5}, is structured around these questions and their corresponding hypotheses. 

\textbf{RQ1: } \textbf{What are the preferences of bilingual readers for annotations in their native language (Tamil/Arabic) versus their second language (English) in visualizations? How do these preferences vary across different chart types and annotation volumes?}
These questions address a gap in the literature on how language influences visualization preferences among bilingual audiences, particularly in the interpretation of chart annotations. Processing information in one's native language generally reduces cognitive effort, enabling faster comprehension and minimizing mental strain. This becomes crucial with higher text density, as second-language processing can overwhelm cognitive resources. Native language annotations also foster stronger emotional and cultural connections, making information more relatable and engaging.  Bilingual readers have been shown to prefer marketing campaigns in their native language as they allow for a more fluid and natural understanding of presented information~\cite{velazquez2017reported}.

\textit{\textbf{H1}: Bilingual readers may show a preference for native language annotations over English annotations, and this preference will increase for increasing volumes of text annotation across all chart types.} 

We also evaluate the effects of finer differences in annotation density and semantic structure on language preferences in our analyses, to assess whether bilingual readers' preferences persist across varying levels of annotation complexity. However, due to the exploratory nature of this study, we did not make specific predictions regarding the strength or direction of the effects for different semantic levels of annotations (L1, L2 for basic facts vs. L3, L4 for high-level context).

\textbf{RQ2:} \textbf{How do bilingual readers use text in their native language vs. second language (English) to effectively interpret and consume chart annotations?} 
This question investigates how language impacts bilingual readers' comprehension of annotations, comparing their accuracy in interpreting visualizations in their native language versus English. Cognitive research suggests that bilingual individuals often exhibit comparable levels of comprehension across languages, especially when the content is familiar or contextually clear~\cite{velazquez2017reported,babino2017like,timmermeister2020no}.

\textit{\textbf{H2}: Bilingual readers will demonstrate increasing levels of chart comprehension for increasing annotation volume in both their native language and English.}

We examine whether participants' self-reported reliance on text annotations influences their ability to interpret annotations across different languages and chart types. Given the exploratory nature of this element of our study, no specific predictions were made about the strength or direction of self-reported reliance on annotations.

\textbf{RQ3: }\textbf{ How do individual variations in linguistic immersion levels influence bilingual readers' preferences for visualization annotation strategies?} This question investigates how various aspects of linguistic immersion, such as fluency, media exposure, internal monologue, technical thought processes, language mixing, and English-medium education, influence bilingual readers' preferences when interpreting annotated visualizations. We explore how different levels of immersion in English versus a native language (Tamil or Arabic) shape preferences for English and non-English annotations. Greater fluency and formal education may reduce cognitive load, facilitating easier processing of complex information in the more fluent language~\cite{babino2017like}. Similarly, frequent media exposure and technical thinking in a particular language could reinforce comfort and familiarity, leading to a stronger preference for annotations in that language~\cite{rosselli2016effect}. On the other hand, frequent language mixing (code-switching) fosters cognitive flexibility, possibly allowing bilinguals to comfortably engage with annotations in either language without a strong preference for one over the other~\cite{velazquez2017reported,napier2005perceptions}.

\textit{\textbf{H3a:} Bilingual readers with higher language fluency, who predominantly consume media, or engage in technical thought in English will exhibit a stronger preference for English annotations of higher volumes across various chart types.}

\textit{\textbf{H3b:} Bilingual readers who frequently engage in language mixing (code-switching) or conduct their internal monologue to an equivalent extent in both languages will demonstrate a more balanced preference for annotations in both English and their native language.}

\textit{\textbf{H3c:} Bilingual readers with extended exposure to formal education in an English-medium will exhibit a stronger preference for annotations in English, across various chart types and annotation volumes.}

By addressing these questions, this study aims to provide insights into the design of effective and inclusive visualizations for bilingual readers, with broader implications for multilingual data communication strategies.


\subsection{Participants}

To determine the sample size, we conducted a power analysis using G*Power~\cite{faul2009statistical}. Based on an SEM model with a WLSMV estimator, an odds ratio of 1.5, and a desired power of 0.8, the required sample size (post-exclusions) was 495 participants per language pair. Accordingly, we recruited 1,096 undergraduate computer science students from two \textit{anonymized universities}, 
in the UAE (\textit{n} = 539, English + Arabic) and India (\textit{n} = 557, English + Tamil). Participants were required to have intermediate or higher fluency in reading, writing, and speaking in both their native language and English. While many participants (71\% across both groups) have intermediate fluency in additional languages, their fluency in these third or fourth languages is often not consistent across reading, writing, and speaking (85\% of multilingual participants), or they have not been formally instructed in those languages (96\% of multilingual participants). This variability would lead to unequal fluency levels if these languages were included in the study. Therefore, we concentrate on the two languages in which participants are equally fluent, ensuring a balanced examination of how language fluency influences interaction with visualizations.

Although currently enrolled in English-medium universities, participants' prior exposure to English varied, with many having studied in Tamil or Arabic-medium schools (74\% of participants) before transitioning to English instruction, ranging from 1 to 15 years (Median Duration of English Education: 2.5 years). This diversity in language experience captures a spectrum of bilingual fluency, providing insights into how varying English exposure influences preferences for visual and textual information. Including this range of experience reflects real-world multilingual contexts and enhances the generalizability of our findings.

All participants had normal color vision. Participants were compensated with university credit~\cite{shen2021student} and not permitted to take the survey multiple times. Study duration averaged roughly 58:36 ($\pm$6:11) minutes. 

We excluded participants who scored below 50\% on any of the screening measures (see Section~\ref{sec:method}), or who provided extremely low-quality responses (i.e., identical answers for all multiple-choice questions or incoherent free-response answers). After exclusions, 507 participants remained in the Arabic condition (i.e., English+Arabic annotations; $\overline{20.35}\pm$0.93 y/o), and 516 participants in the Tamil condition (English+Tamil annotations; $\overline{19.98}\pm$1.56 y/o). 

\subsection{Pilot Testing}

Before conducting the main study, we ran a pilot test\footnote{See Supplemental Material for demographic details and responses.} with a smaller group of participants (10 for each language pair) to refine our survey design and identify potential issues with task clarity and response methods. This led to three modifications in our study mechanism:

(i) During the pilot, participants were initially presented with survey questions in both English and their native language, depending on the language of the charts they were viewing. However, most participants (18/20) found switching between languages during the survey to be disruptive and expressed a preference for consistency in the language of the instructions. This feedback led us to keep all survey instructions in English to provide a smoother experience and maintain consistency, while still allowing participants to engage with charts annotated in both their native language and English. \anj{This decision aligns with participants' exposure to mixed-language environments in digital spaces, where it is common for controls and interfaces to be in English while content, such as posts, are in the native language. For example, on social media—a primary source of exposure cited in participant demographics—controls often remain in English, even when posts are in Tamil or Arabic. News consumption, the second most common source of visualization exposure reported by participants, typically involves content fully localized in one language, such as Tamil or Arabic. However, our study focused on presenting charts as standalone stimuli, akin to how digital visualizations are often consumed in translated form without accompanying articles. We acknowledge that if the charts were presented alongside full articles, consistent language across the article and visualization would likely be expected.}

(ii) In the pilot, participants were given the option to provide their free response answers in either English or their native language (Tamil/Arabic). However, all participants chose to respond in English, citing difficulties with typing in their native language due to unfamiliar keyboard layouts and lack of practice with non-Latin scripts, making the process slow and cumbersome. Based on this feedback, we restricted free-response answers to English in the main study to maintain consistency and ease of expression. While this limited participants' flexibility to use their native language, it minimized typing difficulties and ensured more coherent and comparable responses across participants.

\anj{(iii) To evaluate participant comprehension, we collected participant-generated conclusions (takeaways) in our pilot as part of the free-response task, following methods outlined by Stokes et al.~\cite{stokes2022striking}. However, significant variability emerged in the depth and clarity of participant responses, particularly given the linguistic diversity of our population. This variability made it challenging to draw consistent comparisons across participants. In response, we replaced participant-generated conclusions with a structured conclusion selection task in the main study. This approach was inspired by the structured task design in Stokes et al.~\cite{stokes2022striking}, which systematically evaluated comprehension by presenting participants with predefined conclusions. We accordingly designed three of the five conclusions to align directly with the chart annotations at various semantic levels~\cite{lundgard2021accessible}, while the remaining two conclusions served as distractors-- plausible yet incorrect interpretations of the data-- to test participants' ability to critically evaluate the information. The conclusions were validated by the translators and annotators who participated in the stimuli design process, to ensure that they accurately reflect the data, align with the annotations, and test the intended comprehension aspects. This modification enhanced reproducibility and cross-group comparability, by reducing linguistic and interpretive variability in the comprehension task.
}


\subsection{Study Procedure}
\label{sec:method}

\begin{figure}[H]
    \centering
    \includegraphics[width=0.9\linewidth]{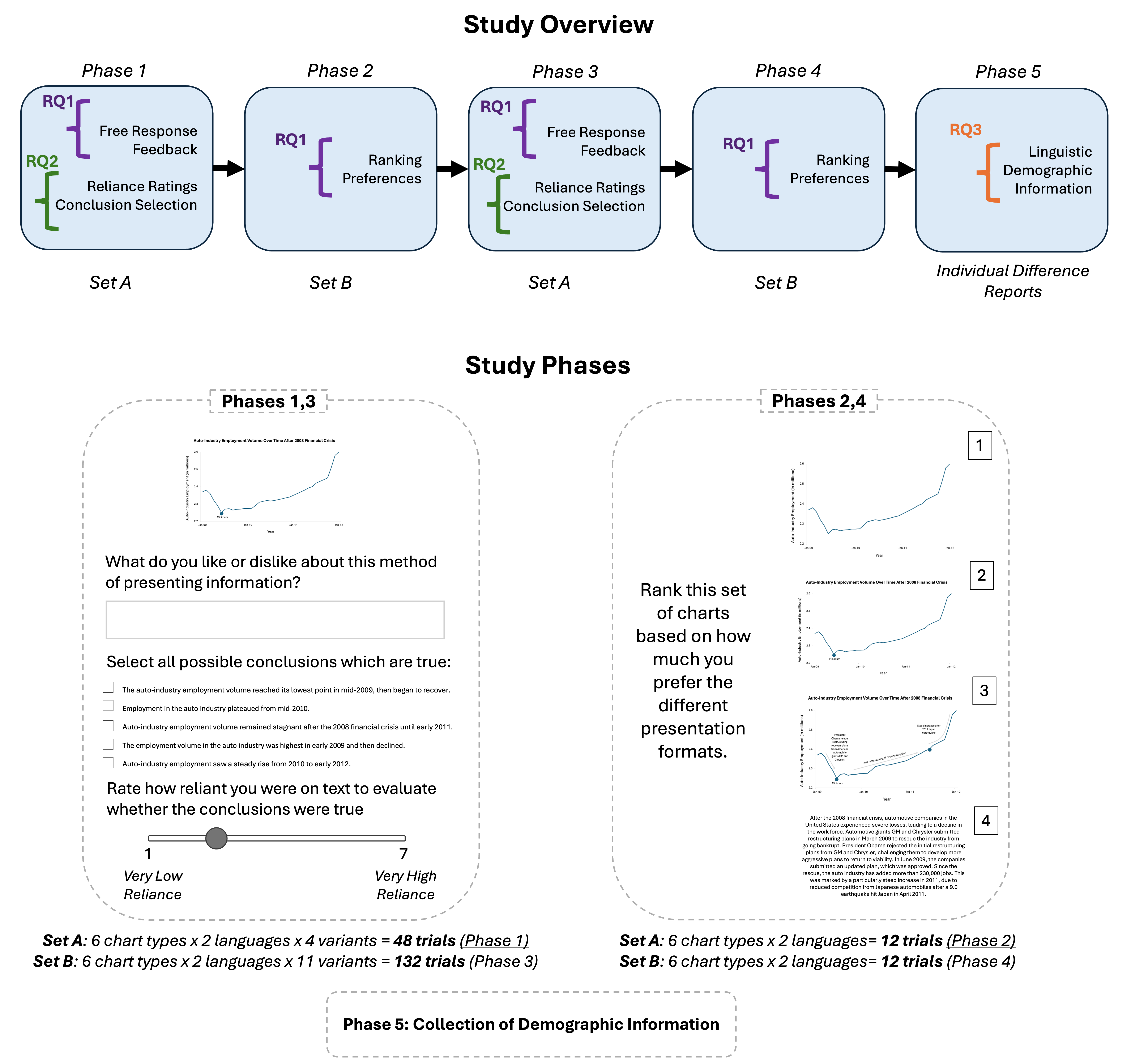}
    \caption{\anj{Participants completed a survey with the five sections. \textit{Top}: Study design summary, depicting the ordering of phases and the respective research questions/hypotheses addressed. \textit{Bottom}: individual viewing of stimuli (phases 1,3), ranking of the full stimuli set (phases 2,4), and demographics (phase 5).}}
    \Description{Study procedure: First, participants state what they like or dislike about a chart, then they select accurate conclusions from a list based on the chart content, and they rate their level of reliance on text while selecting conclusions. This is done for each individual chart belonging to a set. Then, participants rank all charts in the set in order of preference. These tasks are repeated for Set A and then Set B for each of the six chart types. Finally, participants report demographic information.}
    \label{fig:survey}
\end{figure}

Participants underwent three screening tests: (i) a basic comprehension check which involved identifying visual elements (e.g., axes, annotations, and chart titles)~\cite{stokes2022striking}, (ii) the short graph literacy test~\cite{garcia2016measuring}, and (iii) a fluency check which involved successfully completing two training trials covering all the survey tasks in both English and their native language.\footnote{\label{suppsu}See Supplemental Material for exact survey questions.} The survey was terminated for those who scored below 50\% over any of these three tests. After this, participants proceeded to the main survey.

\anj{Participants then completed the main survey, which comprised five sections, as shown in Fig.~\ref{fig:survey}, systematically designed to address the research questions and hypotheses (Table~\ref{tab:rqhmap}). Study trials were completed over one data shape for each chart type, such that at least 365 participants performed tasks for each unique data shape. Below, we outline the study tasks for each phase:}

\textbf{Phase 1 (48 trials):} Using stimuli from Set A (Fig.\ref{fig:seta}) for each of the six chart types, participants completed eight trials (four in English and four in their native language). Each trial consisted of two sub-tasks. First, participants provided a free-response answer (in English) explaining what they liked or disliked about the stimulus shown. \anj{Then, they were presented with five example conclusions about the charts, which were displayed in the same language as the chart annotation as a comprehension check. Participants were instructed to \textit{select the true statements} in this \textit{conclusion selection task}. These conclusions were designed based on the semantic levels (L1–L4) outlined by Lundgard and Satyanarayan~\cite{lundgard2021accessible}, with three directly linked to chart annotations and two serving as distractors. The task structure intentionally aligned with the annotations to isolate their impact on comprehension accuracy.} Finally, participants rated their \textit{``reliance on textual (1) versus visual (7) information''} as a \textit{reliance rating task } when making their selections, following the method used by Stokes et al.~\cite{stokes2022striking}.

\textbf{Phase 2 (12 trials):} For each chart type, all stimuli from Set A used in Phase 1 were displayed simultaneously on the screen. Participants completed two trials per chart type (one in English and one in their native language), across six chart types. In each trial, participants were asked to \textit{rank the set of stimuli based on their overall preference}.

\textbf{Phase 3 (132 trials):} Phase 1 was replicated using stimuli from Set B (Fig.~\ref{fig:setb}), with 11 trials completed in English and 11 in the participants' native language for each of the six chart types.

\textbf{Phase 4 (12 trials):} Phase 2 was replicated using stimuli from Set B, with one trial completed in English and one in the participants' native language for each chart type.

\textbf{Phase 5: } Participants completed the demographic section, where they reported their age, current education level, visualization familiarity \textbf{(VF)}, information representation preference \textbf{(RP)}, and their level of linguistic immersion across various dimensions (see Table~\ref{tab:2}).

\vspace{-6mm}

\section{Quantitative Results}
\label{sec:6}

We utilized structural equation modeling (SEM) to capture the relationships between chart type, annotation level, and participant responses across different languages. Using R (v. 4.3.3), we ran several iterations of multivariate SEM (Structural Equation Modelling)~\cite{joreskog1973analysis} with WLSMV estimator and calculated Cohen's $f^2$ effect size~\cite{selya2012practical}, preserving 0.8 statistical power. Unique SEM iterations were conducted for each language of annotation (English, Tamil, and Arabic), and responses for Set A phases were analyzed separately from those for Set B. Below, we have reported the results for the models that demonstrated the best fit.

In the model, we initially considered responses from Phases 1-4 across both Set A and Set B, we constructed the primary latent independent variable, \textit{(Chart type x Annotation level)}, to account for how these two factors interacted to influence participants' preferences and comprehension. This allowed us to examine how different chart designs and annotation densities shaped responses, while controlling for the effects of language. Dependent variables included ranks assigned to stimuli and conclusion selection accuracy.

\begin{table}[H]
    \centering
    \scriptsize
    \resizebox{0.9\columnwidth}{!}{%
    \begin{tabular}{p{0.2\columnwidth}p{0.7\columnwidth}}
    \toprule
        \textbf{SEM Moderator Variables (Items)} & \textbf{Constituent Variables} \\
        \midrule
        Fluency (6)    &   Reading Fluency ($\oplus$) in (English/Tamil/Arabic), Writing Fluency ($\oplus$) in (English/Tamil/Arabic), Speaking Fluency ($\oplus$) in (English/Tamil/Arabic)\\\midrule
        Media Exposure (4)   &   Language in which you consume ($\triangle$): Television/Radio/Books for fun/Social Media \\\midrule
        Internal Monologue (4)   &  Language in which you do the following activities ($\triangle$): Thinking/Dreaming/Emoting/Talking to Self \\\midrule
        Technical Thought (6)   &   Language in which you do the following activities ($\triangle$): Mental Arithmetic/Remembering Information/Searching for Information/Discovering Information/Reading Technical Content/Forming Hypotheses
        \\\midrule
        Language Mixing (4)   &   Frequency with which you do the following activities ($\bullet$): Switching between languages: talking to friends/talking to family/internal monologue/technical thought\\\midrule
        English Education Duration (1)   &  Number of years educated in English-medium \\
        \bottomrule        
    \end{tabular}%
    }
    \caption{\anj{Moderator Variables used in SEM, computed as latent variable constructs over Demographic Information collected in Study \textbf{Phase 5}. The individual items within each construct are rated on a scale from 1 (``English''$\triangle$/``Very Rare''$\bullet$/``Very Low''$\oplus$) to 7 (``Tamil (or) Arabic''$\triangle$/``Very Frequent''$\bullet$/``Very High''$\oplus$). English Education Duration (number of years) is examined a direct measure.} }
    \label{tab:2}
\end{table}

\begin{figure}[H]
\centering
    \includegraphics[width=\columnwidth]{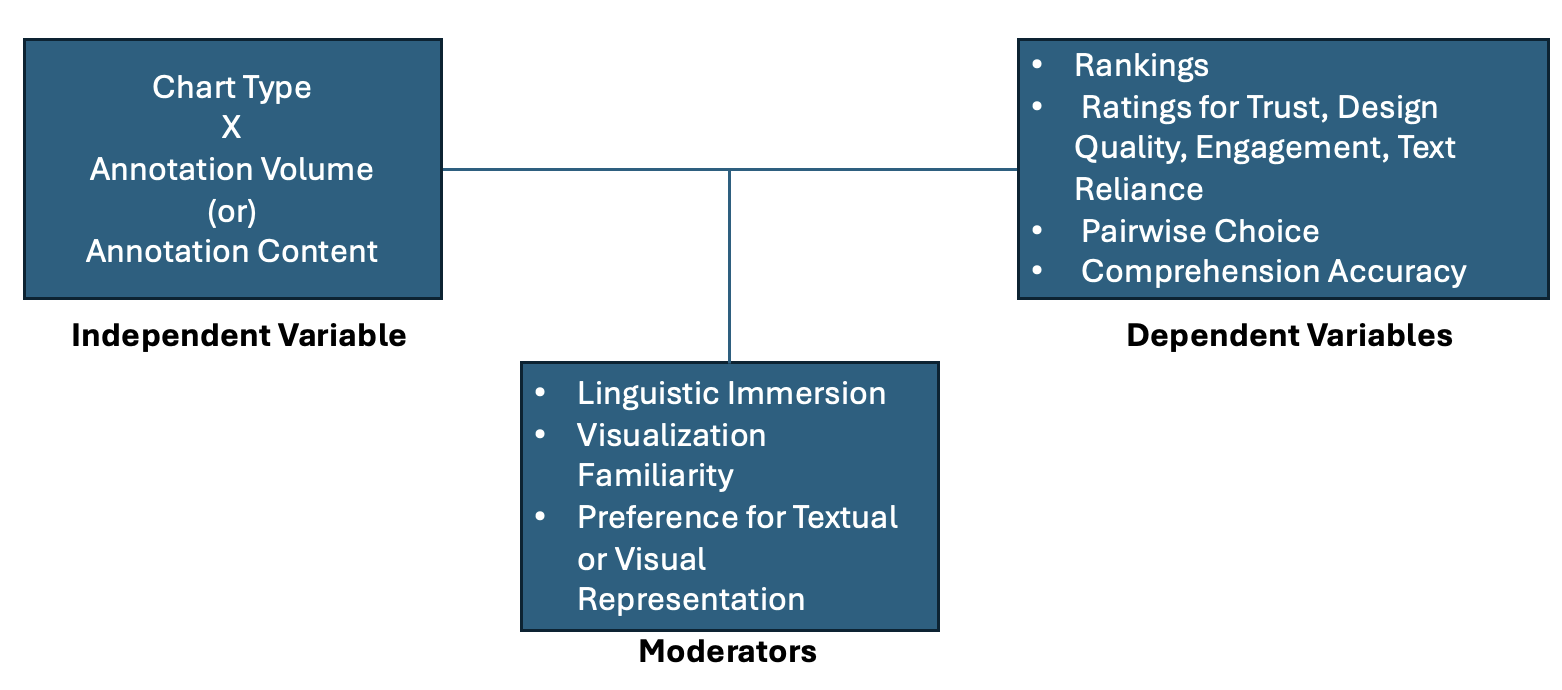}
    \caption{
    SEM analysis structure. SEM is run separately across each annotation language tested (\eng{English}, \ta{Tamil}, \ab{Arabic}). We use latent constructs for the independent variable as Chart x Annotation Level (volume variants for Set A and semantic content for Set B stimuli), and for the linguistic immersion moderators. All other variables are observed measures. Link color and thickness denote the value range for interaction coefficients ($\beta$) corresponding to the links, averaged across all three languages tested.
    }
    \Description{SEM model: We use latent constructs for the independent variable as Chart x Annotation Level (volume variants for Set A and semantic content for Set B stimuli), and for the linguistic immersion moderators. Text reliance is an observed measure also used as a moderator. Dependent variables are ranks assigned to stimuli and conclusion selection accuracy.}
    \label{fig:SEM}
\end{figure}

\begin{figure*}[!ht]
\centering
    \includegraphics[width=0.8\textwidth]{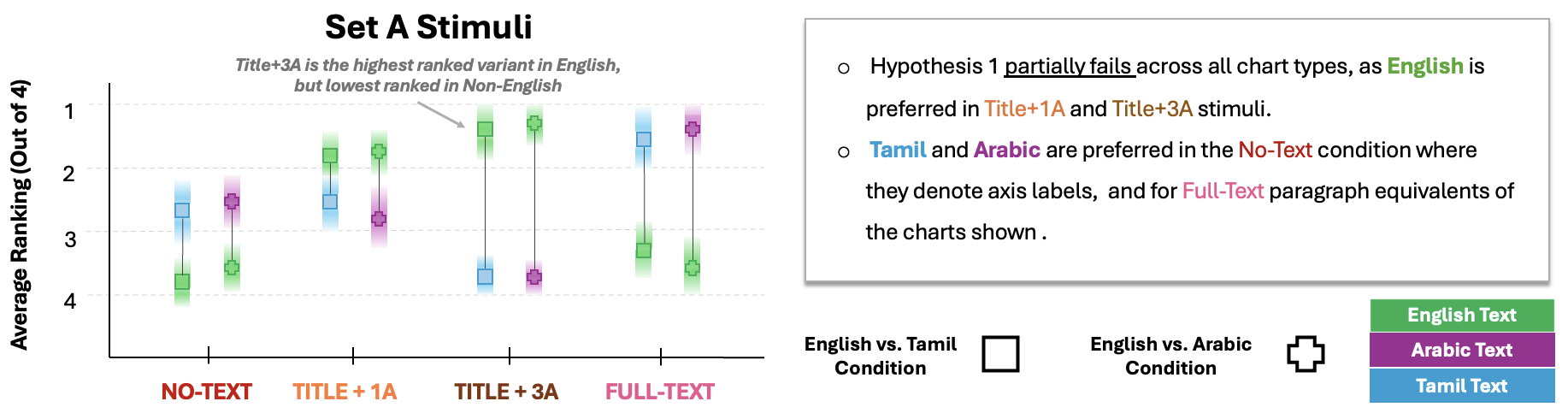} 
    \caption{
    Plot showing a summary of ranking patterns from Phase 2, done for stimuli belonging to Set A. We show average rank across different text variants, collapsed across all chart types tested. The color of the points represents the language of annotation, and the shape represents the bilingual condition tested (\eng{English} vs. \ta{Tamil} or \eng{English} vs. \ab{Arabic}). 
    Gradient bands represent uncertainty associated with the observed rankings in terms of standard deviations.
    }
    \Description{Summary of ranking patterns for Set A.  Hypothesis 1 partially fails across all chart types, as English is preferred in Title+1A and Title+3A stimuli. Tamil and Arabic are preferred in the No-Text condition, where they denote axis labels,  and for Full-Text paragraph equivalents of the charts shown.}
    \label{fig:ranksa}
\end{figure*}

We then incorporated text reliance ratings from Phases 1 and 3, as well as demographic responses from Phase 5, which measured linguistic immersion (detailed in Table~\ref{tab:2}), as moderators to explore how they influenced the interaction between language and visualization design. The remaining demographic data \anj{(participants' age, gender, familiarity with the chart types presented, visualization literacy, technological literacy)} collected are used as control variables in our modeling.

We illustrate the overall structure of our SEM approach in Figure~\ref{fig:SEM}. We specify model paths connecting every possible unique combination of <independent--moderator--dependent> paths in our model. Below, we discuss the normalized regression results from our modeling.\footnote{\label{foot:sig}All results reported in the main paper were significant, with $p<0.05$.}


\subsection{RQ1: Language Preferences Across Annotation Levels}

Our first research question aimed to explore how bilingual readers' preferences for annotations in their native language (Tamil/Arabic) versus their second language (English) differ across varying chart types and annotation volumes. Specifically, we sought to understand whether participants would gravitate towards annotations in their native language, especially as the amount of text increased, or if their second language (English) would prove preferable due to familiarity with its usage in scientific and academic contexts. To test \textbf{H1}, which posited that bilingual readers may prefer native language annotations over English annotations, we examined how \textbf{language}, \textbf{annotation density}, and\textbf{ semantic content} predicted participants' rankings and choices in our model.

\subsubsection{How are language preferences driven by annotation density? (Fig.~\ref{fig:ranksa})}

\leavevmode

\textbf{Heavily annotated \eng{\textbf{English}} charts are preferred.} Participants consistently ranked \eng{English-annotated charts} (English$_{Tamil}$: $1.41\pm0.32$; English$_{Arabic}$: $1.74\pm0.42$) with higher annotation density (\threeA{title+3A}) as their top choice ($\overline{\chi^2} = 148.53$; $\overline{\beta} = 0.69$, $\overline{f^2} = 0.38$) across all chart types \footnote{Unless otherwise specified, all reported values represent averages across the six chart types tested.}, particularly for \textbf{scatterplots} and \textbf{heatmaps}, where 84\% ranked them highest ($\overline{\beta} = 0.82$, $\overline{f^2} = 0.63$), compared to 59\% for other chart types ($\overline{\beta} = 0.59$, $\overline{f^2} = 0.41$). This aligns with prior findings that English annotations with rich semantic content (L3/L4) enhance clarity~\cite{stokes2022striking,stokes2023role}.\footnote{Here, `Tamil' and `Arabic' refer to two native-language conditions tested in our study. When an overline is present (e.g., $\overline{\chi^2}$, $\overline{\beta}$), the reported regression values represent averages across both these conditions.}

This strong preference for English may stem from its dominance in academic and scientific contexts, where participants are accustomed to processing data visualizations. English is often associated with precision and clarity, particularly for technical content. Additionally, the cognitive effort required to interpret complex annotations in a non-native language may have contributed to participants favoring English for high-density annotations.

\begin{figure}[H]
\centering
    \includegraphics[width=0.4\textwidth]{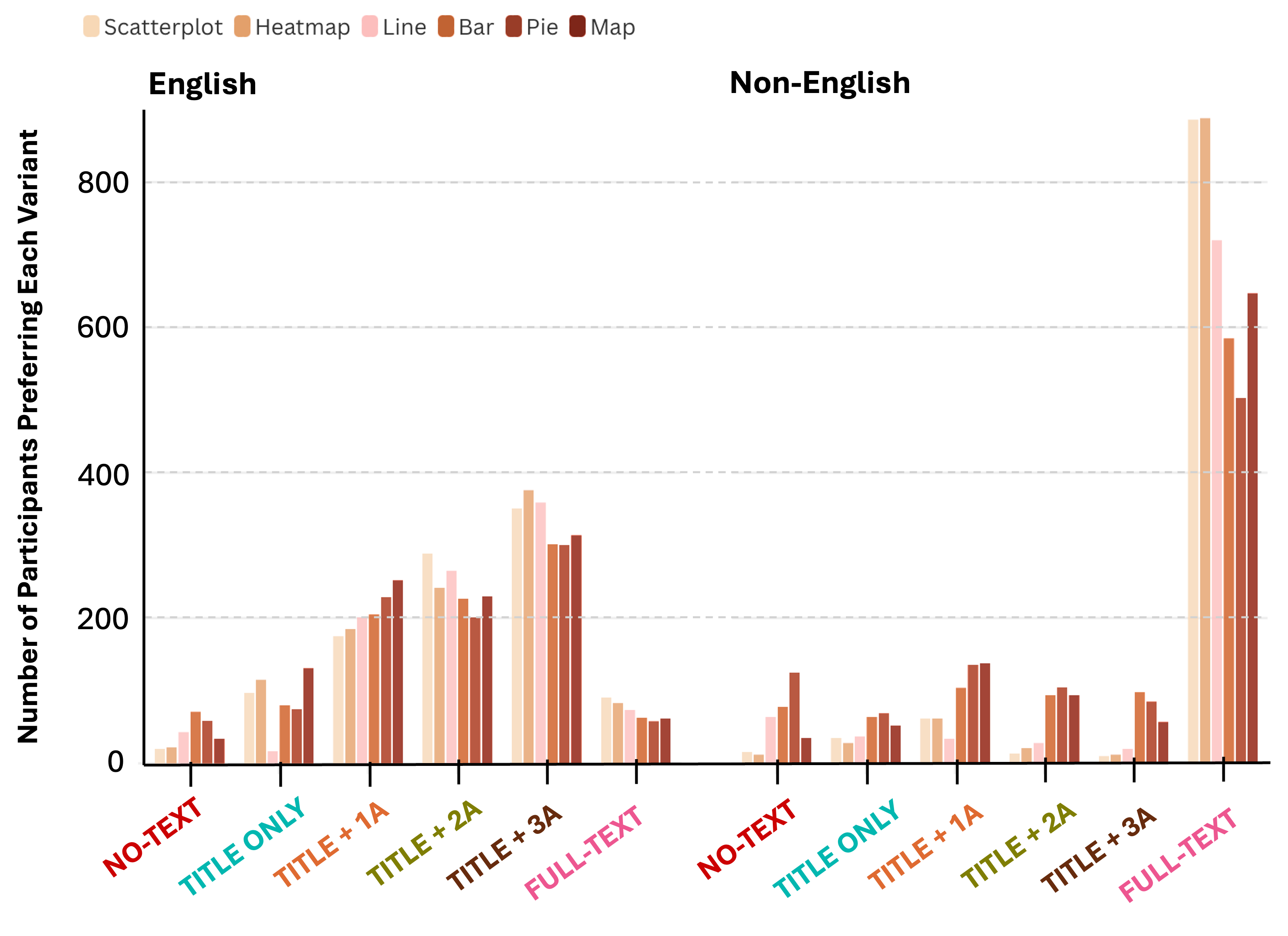}
    \caption{Number of participants who ranked each annotation variant first (across Sets A and B) by chart type and language. \eng{English} annotations: \threeA{Title+3A} was most preferred, especially for scatterplots and heatmaps, while \full{Full-Text} was least favored. \ta{Tamil}/\ab{Arabic} annotations: \full{Full-Text} was strongly preferred, particularly for scatterplots and heatmaps, indicating a preference for narrative-style text in native languages. These trends highlight differences in how bilingual participants engage with English vs. Non-English annotations based on annotation density and chart type.
    }
     \Description{The figure shows the number of participants who preferred each annotation variant (No-Text, Title-Only, Title+1A, Title+2A, Title+3A, Full-Text) across different chart types (Scatterplot, Heatmap, Line, Bar, Pie, and Map) for both English and Non-English languages. For English annotations, Title+3A emerged as the most preferred variant across all chart types, with Scatterplots and Heatmaps showing particularly high participant counts. Conversely, Full-Text annotations in English were the least preferred, indicating that participants favored charts with on-chart text annotations over narrative-style paragraphs. For Non-English annotations, Full-Text was strongly favored across all chart types, particularly for complex charts like Scatterplots and Heatmaps. This suggests a preference for narrative-based text when participants process visualizations in their native language. Overall, the trends highlight a divergence in how bilingual participants engage with English versus Non-English annotations, depending on annotation density and chart type.
}
    \label{prefcounts}
\end{figure}

\begin{figure*}[!ht]
\centering
    \includegraphics[width=0.95\textwidth]{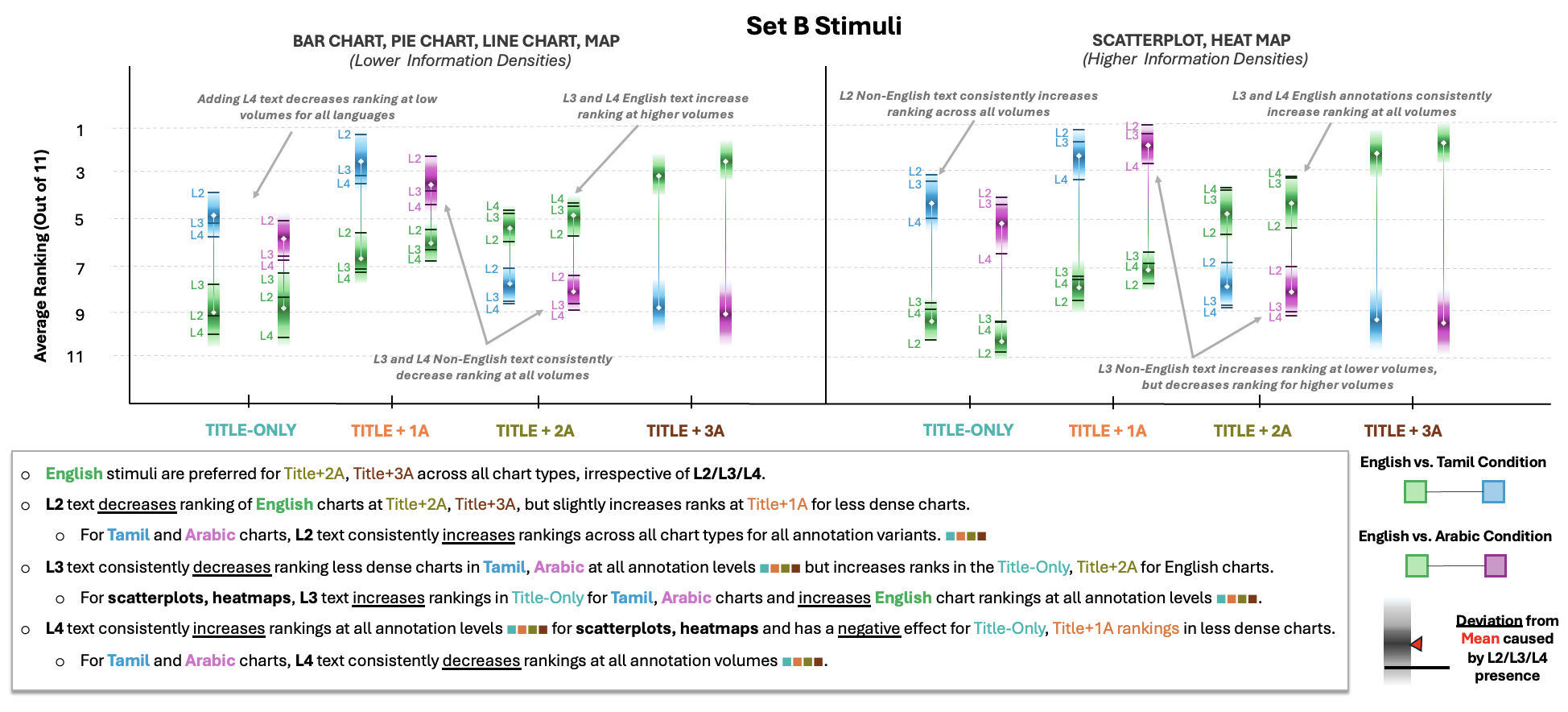}
    \caption{
    Small multiple plots showing a summary of ranking patterns from Phase 4, for stimuli in Set B. We show average rank per low information-dense (left panel) and high information-dense (right panel) chart types across different text variants. The color of the points represents the language of annotation for the two conditions tested (\eng{English} vs. \ta{Tamil} or \eng{English} vs. \ab{Arabic}). Key comparisons are detailed at the bottom. Gradient bands represent uncertainty associated with the observed rankings in terms of standard deviations; dark lines represent the deviation in ranking from the mean caused by the presence of an L2/L3/L4 annotation.
    }
     \Description{Summary of ranking patterns for Set B.   English stimuli are preferred for Title+2A, Title+3A across all chart types, irrespective of L2/L3/L4. L2 text decreases ranking of English charts at Title+2A, Title+3A, but slightly increases ranks at Title+1A for less dense charts. For Tamil and Arabic charts, L2 text consistently increases rankings across all chart types for all annotation variants. L3 text consistently decreases ranking less dense charts in Tamil, Arabic at all annotation levels but increases ranks in the Title-Only, Title+2A for English charts. For scatterplots, heatmaps, L3 text increases rankings in Title-Only for Tamil, Arabic charts and increases English chart rankings at all annotation levels. L4 text consistently increases rankings at all annotation levels for scatterplots, heatmaps and has a negative effect for Title-Only, Title+1A rankings in less dense charts. For Tamil and Arabic charts, L4 text consistently decreases rankings at all annotation volumes.
}
    \label{fig:ranksb}
\end{figure*}

\textbf{Minimal annotation levels are preferred in \ta{Tamil} and \ab{Arabic} charts.} 
In contrast, participants favored charts with minimal native-language annotations (\noT{no-text}) in non-English stimuli ($\overline{\chi^2} = 162.37$, $\overline{\beta} = 0.76$, $\overline{f^2} = 0.44$), with only axis labels present. Full-text summaries were also favored in \ta{Tamil} ($1.29\pm0.15$) and \ab{Arabic} ($1.43\pm0.28$), particularly for \textbf{scatterplots} and \textbf{heatmaps}, where 87\% ranked \full{full-text} highest, while 68\% preferred \noT{no-text} as the second-best option. Dense non-English annotations (\threeA{title+3A}) were generally ranked lower, with 54\% of participants selecting them as the least preferred for \textbf{scatterplots} and \textbf{heatmaps}, compared to 31\% for other chart types. This suggests a preference for simpler, more direct textual information in native languages when dealing with high-density visualizations.

These preferences may be linked to how participants engage with text based on location and density. Full-text summaries provide a narrative format that is cognitively demanding in a second language, leading participants to favor their native language for detailed explanations. In contrast, smaller embedded annotations (e.g., \oneA{title+1A}, \threeA{title+3A}) may be easier to process in English, especially for participants accustomed to consuming technical content in English.

Figure~\ref{prefcounts} further illustrates participant preferences across annotation levels and chart types, highlighting a preference for denser English annotations and minimal or full-text annotations in Tamil/Arabic. These trends diminish for less information-dense charts like bar and pie charts, where preferences are more balanced.

\subsubsection{How does the semantic content of annotations shape language preferences? (Fig.~\ref{fig:ranksb})} 
\label{s612}

\leavevmode

\eng{English} stimuli with high-level semantic content are preferred. Participants ranked \textbf{L3/L4} \eng{English} annotations higher than L2 across all annotation levels $\tO{\blacksquare}\oneA{\blacksquare}\twoA{\blacksquare}\threeA{\blacksquare}$ ($\overline{\beta} = 0.71$, $\overline{f^2} = 0.36$) for \textbf{scatterplots} and \textbf{heatmaps}, where contextual information helped interpret complex data (Tamil (L3/L4): $9.18\pm0.42$; Arabic (L3/L4): $8.89\pm0.26$). However, \textbf{L4} annotations decreased preference for low-density charts in the \tO{title-only} and \oneA{title+1A} conditions (Fig.~\ref{fig:ranksb}, left panel).

For \textbf{scatterplots} and \textbf{heatmaps}, \textbf{L1/L2} \eng{English} annotations were ranked lower overall (English$_{Tamil}$ (L1/L2): $\overline{\triangle } \uparrow 0.78$; English$_{Arabic}$ (L1/L2): $\overline{\triangle } \uparrow 1.03$), indicating participants found higher-level content more useful in their second language. In contrast, for low-density charts, \textbf{L1/L2} text decreased rankings at higher annotation volumes (\twoA{title+2A}, \threeA{title+3A}) but increased rankings at lower annotation volumes (\tO{title-only}, \oneA{title+1A}).

Despite these differences, higher annotation levels (\twoA{title+2A}, \threeA{title+3A}) preserved the effect of L2/L3/L4 content across \eng{English}, \ta{Tamil}, and \ab{Arabic}. This suggests that when annotation density is high, participants rely on overall information availability rather than semantic depth, as comprehensive text provides sufficient context for interpretation.

\begin{figure*}[!ht]
    \centering
    \includegraphics[width=0.9\textwidth]{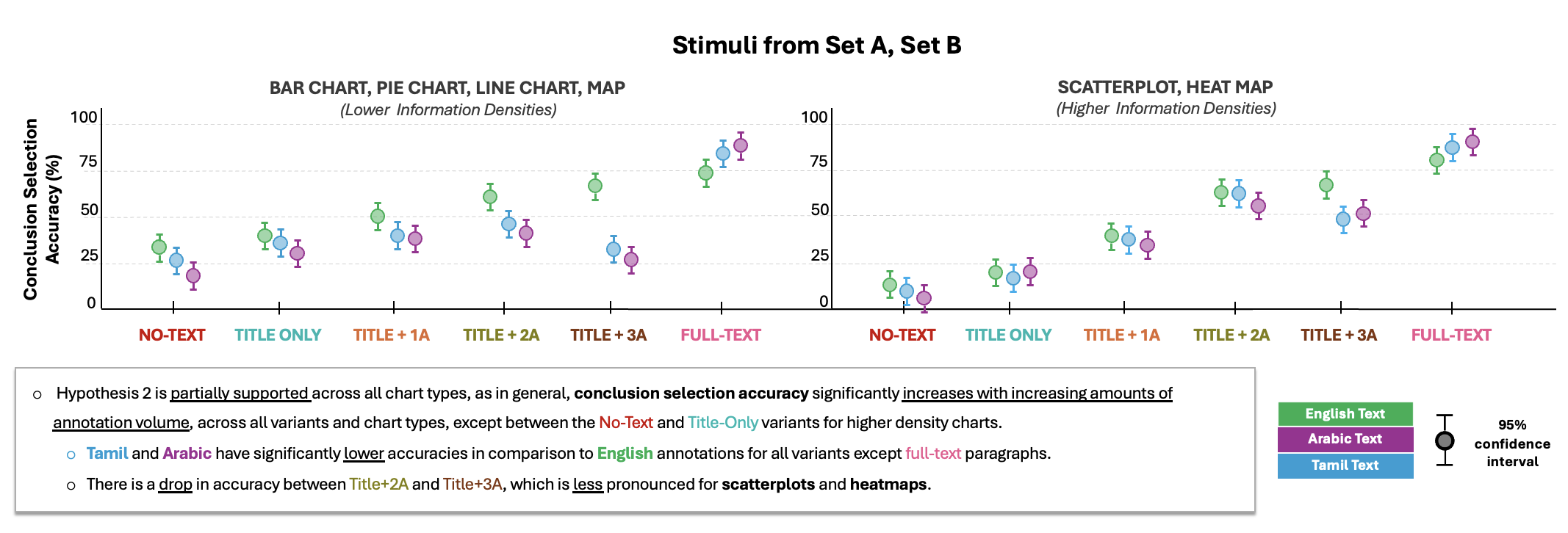}
    \caption{Small multiple plots showing a summary of conclusion selection accuracies from Phases 1 and 3, for stimuli in both Set A and Set B. We show average accuracy per low information-dense (left panel) and high information-dense (right panel) chart types across different text variants. The color of the points represents the language of annotation for the two conditions tested (\eng{English} vs. \ta{Tamil} or \eng{English} vs. \ab{Arabic}). The accuracies for stimuli annotated in \eng{English} are combined, as they display negligible variation ($<5\%$) between participant groups. Key comparisons are detailed at the bottom. Error bars represent uncertainty associated with the observed accuracies with a 95\% confidence interval.}
    \Description{ Hypothesis 2 is partially supported across all chart types, as in general, conclusion selection accuracy significantly increases with increasing amounts of annotation volume, across all variants and chart types, except between the No-Text and Title-Only variants for higher density charts. Tamil and Arabic have significantly lower accuracies in comparison to English annotations for all variants except full-text paragraphs. There is a drop in accuracy between Title+2A and Title+3A, which is less pronounced for scatterplots and heatmaps.}
    \label{fig:6.2}
\end{figure*}

\textbf{\ta{Tamil} and \ab{Arabic} charts with basic semantic content (L1/L2) are preferred.} Participants preferred basic, factual annotations (\textbf{L1/L2}) over more complex, interpretive content (L3/L4), particularly in \tO{title-only} and \oneA{title+1A} conditions ($\overline{\beta} = 0.75$, $\overline{f^2} = 0.39$). Adding \textbf{L4} content lowered rankings across annotation levels, $\tO{\blacksquare}\oneA{\blacksquare}\twoA{\blacksquare}\threeA{\blacksquare}$ (\ta{Tamil}: $3.27\pm0.20$, $\overline{\triangle} L3,L4 \uparrow 1.17$; \ab{Arabic}: $3.16\pm0.18$, $\overline{\triangle} L3,L4 \uparrow 1.22$), suggesting a preference for simpler, more direct information when reading in their native language. Conversely, L3 was seen to increase rankings for the \tO{title-only} and \oneA{title+1A} variants in the case of \textbf{scatterplots} and \textbf{heatmaps}. This suggests that participants found L4 annotations in their native language less useful or perhaps more cumbersome to process, while L3 annotations helped focus analysis, especially when combined with lower-level factual information (English$_{Tamil}$: $7.06\pm0.32$, $\overline{\triangle} L1,L2 \uparrow 0.74$; English$_{Arabic}$: $7.25\pm0.15$, $\overline{\triangle} L1,L2 \uparrow 0.58$). 

\textit{Hence, while participants did not consistently prefer native language annotations across all chart types and annotation volumes, they did favor non-English full-text annotations with higher information density, offering partial support for \textbf{H1}.}


\subsection{RQ2: Comprehension Accuracy and Information Reliance (Fig.~\ref{fig:6.2})}

We sought to determine whether bilingual participants demonstrate similar levels of accuracy when interpreting visualizations across different languages and how the presence and volume of textual annotations impact their ability to correctly identify key conclusions, i.e., takeaways, from the charts. To test H2, which focused on the role of annotation complexity and chart type in driving participants' comprehension across languages, in our model, we focused on two key measures: conclusion selection accuracy and self-reported levels of reliance on textual information.

\textbf{Higher annotation levels improve conclusion selection accuracy in \eng{English}, but have mixed effects in \ta{Tamil} and \ab{Arabic}.} Participants showed higher accuracy with more detailed \eng{English} annotations (\twoA{title+2A}, \threeA{title+3A} (Mean: 0.63, SD: 0.12; $\overline{\beta} = 0.71$, $\overline{f^2} = 0.36$), while \noT{no-text} and \tO{title-only} variants performed worst (Mean: 0.24, SD: 0.07). \full{Full-text} paragraphs yielded the highest accuracy overall. Accuracy was lowest for \textbf{scatterplots} and \textbf{heatmaps} without detailed annotations $\noT{\blacksquare} \tO{\blacksquare} \oneA{\blacksquare}$ ($\overline{\triangle} \downarrow 0.23$), suggesting these charts require more textual support to elicit conclusions.

For non-English annotations ($\overline{\beta} = 0.75$, $\overline{f^2} = 0.39$), accuracy generally increased with annotation volume but dropped from \twoA{title+2A} to \threeA{title+3A}. This may be due to \textbf{L4} content shifting focus away from chart interpretation, increasing cognitive load in \threeA{title+3A} (Section~\ref{s612}). This shift in focus likely makes it more challenging for users to process the data effectively, contributing to the observed decrease in ranking. However, semantic levels did not have a statistically significant effect across all chart types ($p>0.05$).

We note that non-English annotations had lower accuracy than \eng{English} for visually annotated charts (\ta{Tamil}: $\overline{\triangle} \downarrow 0.09$, \ab{Arabic}: $\overline{\triangle} \downarrow 0.16$), especially for low-density charts ($\overline{\triangle} \downarrow 0.21$). Full-text variants, however, performed best in non-English (\ta{Tamil}: Mean 0.81, SD 0.06; \ab{Arabic}: Mean 0.86, SD 0.05), likely due to reduced cognitive demand when processing narrative text in their native language. \full{Full-text} paragraphs provide a linear narrative, which may be more intuitive when presented in their native language. In contrast, differences in reading direction or stroke density of the annotations might make it harder to integrate textual information with the visual data in non-English languages.

\textit{Hence, these results partially support \textbf{H2}, as participants demonstrated increasing levels of comprehension across languages when increasing annotation volumes were present.}

\textbf{Text reliance increases with chart information density, but does not moderate comprehension accuracy across languages.} Participants relied more on text for high-density charts like \textbf{scatterplots} and \textbf{heatmaps}, regardless of annotation complexity. For \eng{English} annotations, text reliance was moderate across all variants (English$_{Tamil}$: $4.16\pm0.53$; English$_{Arabic}$: $4.53\pm0.34$), but highest for \textbf{scatterplots} and \textbf{heatmaps} ($\overline{\triangle} \uparrow 1.33$). This pattern held for most non-English annotations $\noT{\blacksquare} \tO{\blacksquare} \oneA{\blacksquare} \twoA{\blacksquare}$, where \textbf{scatterplots} and \textbf{heatmaps} again required the most reliance on text (\ta{Tamil}: $5.41\pm1.04$; \ab{Arabic}: $5.23\pm0.84$), while \textbf{bar} and \textbf{pie} charts showed lower reliance (\ta{Tamil}: $2.58\pm0.61$; \ab{Arabic}: $2.47\pm0.39$). Although detailed annotations (\threeA{title+3A}, \full{full-text}) further increased text reliance for complex charts, this did not significantly moderate comprehension accuracy ($p>0.05$).

Overall, the presence of detailed annotations improved comprehension across all languages, particularly in cognitively demanding charts (scatterplots and heatmaps). Our results suggest that the availability of sufficient annotation was more critical to comprehension than the semantic content or extent to which participants relied on it. 


\subsection{RQ3: Effects of Individual Differences on Language Preferences}
\label{sec63}

\begin{figure*}[!ht]
\centering
    \includegraphics[width=\textwidth]{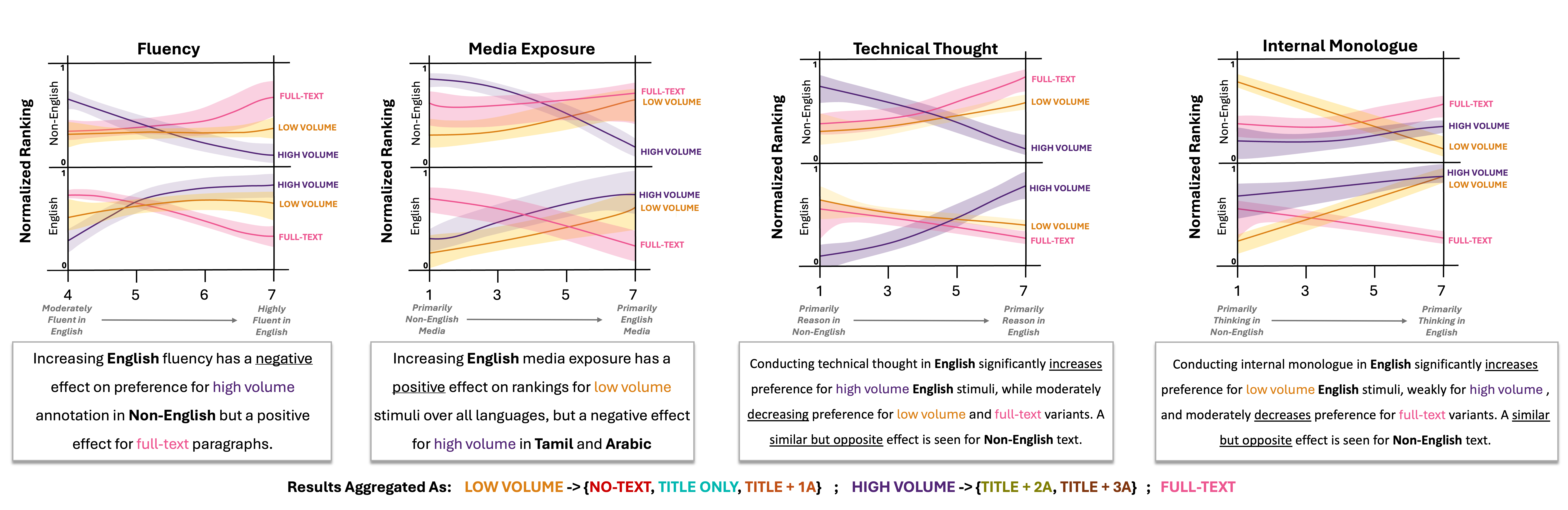}
    \caption{Small-multiple multi-line charts with 95\% confidence band, showing the impact of linguistic immersion moderators on ranking patterns across both Set A and Set B. Each small multiple represents a moderating variable, with the x-axis displaying increasing values that denote progressively greater usage of English (and simultaneous decline of Non-English) for the constituent activities of each respective moderator. Min-max normalization is applied to allow comparison of the moderation effects on both Set A and Set B simultaneously (0: low rank, 1: high rank, where higher ranked stimuli are preferred by participants). 
    Annotations are organized into three groups: \lav{low volume} (no-text, title only, and title+1A), \hav{high volume} (title+2A and title+3A), and \full{full-text} stimuli. 
    The bottom row of charts denotes the effects on \textbf{English} and the top row denotes effects aggregated over \textbf{Tamil+Arabic} annotations.}
    \Description{Increasing English fluency has a negative effect on preference for high volume annotation in Non-English but a positive effect for full-text paragraphs. Increasing English media exposure has a positive effect on rankings for low volume stimuli over all languages, but a negative effect for high volume in Tamil and Arabic. Conducting technical thought in English significantly increases preference for high volume English stimuli, while moderately decreasing preference for low volume and full-text variants. A similar but opposite effect is seen for Non-English text. Conducting internal monologue primarily in English has the strongest effects on low-volume stimuli for both English and Non-English charts, but with opposite directionalities. When monologuing is done equally in both languages, the assigned ranks are comparable.}
    \label{fig:6.3}
\end{figure*}

In exploring how individual differences shape bilingual readers' preferences for visualization annotations, we aimed to understand how varying levels of exposure to English or native languages (Tamil/Arabic) influence participants’ preferences for English or non-English annotations in data visualizations. To test H3a--c, we specifically examined how facets of linguistic immersion moderated ranking and comprehension accuracy in our model. This approach sheds light on the cognitive and experiential factors that drive language preferences in bilingual contexts. The results below provide insights into how these individual differences influenced annotation preferences and comprehension accuracy across different chart types and annotation volumes; we summarize the significant effects seen in Fig.~\ref{fig:6.3}).

\begin{table}[H]
    \centering
    \scriptsize
    \resizebox{\columnwidth}{!}{%
    \begin{tabular}{lcc}
    \toprule
         & \textbf{English--Tamil} &  \textbf{English--Arabic} \\
         \midrule
        \textbf{Number of Participants} & 516 & 507 \\
        \textbf{Age (years, mean (SD))} & 19.98 (1.44) & 20.35 (0.93)  \\
        \textbf{Gender (M/F/)} & 277/239 & 280/227  \\
        \textbf{Years of English-medium Education (mean (SD))} & 3.57 (1.59) & 3.85 (1.50)  \\
        \textbf{Media Exposure} & 4.83 (1.43) & 4.81 (1.53) \\
        \textbf{Internal Monologue} & 4.68 (1.35) & 4.73 (1.24) \\
        \textbf{Technical Thought} & 4.16 (1.53) & 4.05 (1.18) \\     
        \textbf{Language Mixing During Technical Thought} & 3.03 (1.40) & 3.05 (1.43)\\     
        \textbf{Language Mixing During Internal Monologue} & 4.94 (1.39) & 5.04 (1.43) \\     
        \textbf{Language Mixing During Conversation} & 6.27 (0.93) & 6.31 (0.88) \\     
        \textbf{Visualization Familiarity} & 4.35 (0.85) & 4.10 (1.30)\\
        \textbf{Technological Literacy} & 4.15 (0.90) & 4.25 (1.05)\\
    \bottomrule
    \end{tabular}%
    }
    \resizebox{\columnwidth}{!}{%
    \begin{tabular}{lcccc}
         & \textbf{English$_{Tamil}$} & \textbf{Tamil} & English$_{Arabic}$&\textbf{Arabic} \\
         \midrule
    \textbf{Reading Fluency} & 5.78 (0.87) & 5.89 (1.05) & 5.80 (0.83) & 5.89 (1.21)\\
    \textbf{Writing Fluency} & 5.73 (1.24) & 5.88 (1.03) & 5.75 (1.17) & 5.87 (1.09)\\
    \textbf{Speaking Fluency} & 6.02 (0.58) & 6.95 (0.62) & 5.96 (0.55) & 6.93 (0.71)\\
    \bottomrule
    \end{tabular}%
    }
    \caption{\anj{Top: Participant demographics across the two language groups. The scales for Media Exposure, Technical Thought, and Internal Monologue 
    are oriented such that 1 = strong preference for English and 7 = strong preference for Tamil/Arabic. For Language Mixing and Familiarity ratings, 1 = very low/rarely and 7 = very high/frequently. Bottom: Fluency ratings for participants for each language tested. High values for fluency (above 5) confirm that participants are highly fluent in both English and their native language.}}
    \label{demographics}
\end{table}

\textbf{Demographic Characteristics:} Table~\ref{demographics} summarizes participant demographics. Both language groups (English–Tamil, English–Arabic) were similar in age (~20 years) with a slight male majority. Participants had 3–4 years of English-medium education and reported high fluency (above 5 on a 7-point scale) in both English and their native language, confirming balanced bilingualism. Preferences for English vs. Tamil/Arabic varied by domain (e.g., media exposure, internal monologue, technical thought) but centered around the midpoint (4 on a 1–7 scale), reflecting mixed language usage. Participants also reported moderate-to-high language mixing in conversation and internal thought. Familiarity with data visualizations and technological literacy (both just above 4 on a 1–7 scale) suggested a moderate baseline proficiency in data interpretation. Overall, the sample consisted of young, balanced bilinguals with strong language skills and familiarity with visual and technological content.

\textbf{Fluency, Media Exposure, and Technical Thought drive preference for English annotations.} Participants with higher English fluency, media exposure, or technical thought in English consistently preferred English annotations and performed better in conclusion selection tasks.

\textbf{\textit{Fluency:}} Greater \textbf{English} fluency correlated with higher rankings for \textbf{English} charts with \hav{high annotation volumes} ($\overline{\beta} = 0.64$, $\overline{f^2} = 0.46$), particularly for \textbf{scatterplots} and \textbf{heatmaps} ($\overline{f^2} = 0.66$). Conclusion selection accuracy increased by 47\% in these cases. Conversely, in \textbf{non-English} charts, only \full{full-text} annotations saw increased preference ($\uparrow 28\%$), while high-density annotations were less favored ($\downarrow 34\%$). \lav{Low annotation volumes} on the other hand, experienced nonsignificant increases for all chart types and annotation languages. 

\textbf{\textit{Media Exposure:}} Participants who consumed \textbf{English} media ranked \lav{low-annotation} \textbf{English} and \textbf{non-English} charts moderately higher ($\uparrow 38\%$, $\overline{\beta} = 0.47$, $\overline{f^2} = 0.49$), with increased conclusion selection accuracy in \textbf{English} ($\uparrow 25\%$, ). \hav{High-annotation} charts showed mixed effects \textbf{English}: $\downarrow 23\%$), and non-English: $\downarrow 36\%$). Strongest effects appeared in \textbf{line charts}, \textbf{maps}, and \textbf{scatterplots} ($\overline{f^2} = 0.54$). 

\textbf{\textit{Technical Thought:} } Those using English for technical tasks strongly preferred high-annotation English charts ($\uparrow 68\%$) and avoided high-annotation non-English charts ($\downarrow 55\%$). Full-text non-English annotations saw a slight increase in ranking ($\uparrow 21\%$), while their English counterparts decreased ($\downarrow 27\%$); \lav{low annotation volumes} behaved similarly to the \full{full-text} variant across all languages, but the ranking changes were not found to be significant.  The preference for high annotation volumes during technical thought likely stems from the nature of tasks like analysis and memory recall, which require detailed, structured information to aid in processing complex data. 

Overall, moderator effects were most notable for charts requiring detailed analysis, such as \textbf{scatterplots} and \textbf{heatmaps}, where participants found \eng{English} annotations more effective for processing complex data. Additionally, conclusion selection accuracy ($\overline{\beta} = 0.75$, $\overline{f^2} = 0.54$) for all variants of \eng{English} annotations was significantly higher (accuracy: $\uparrow 47\%$). Frequent English engagement in media and technical contexts likely reduces cognitive load, making English annotations easier to process, especially for analytical tasks. The accessibility of academic and technical terms in English may further enhance comprehension and accuracy in these cases.

\textit{Hence, increased linguistic immersion in English (through fluency, media exposure, and technical thought) positively correlates with a preference for higher volumes of English annotation, supporting \textbf{H3a}.}

\textbf{Internal Monologue drives language preferences at low annotation volumes.} 

Participants who primarily thought in \textbf{English} strongly preferred \lav{low annotation volume} \textbf{English} charts ($\uparrow 64\%$, $\overline{\beta} = 0.77$, $\overline{f^2} = 0.52$) and were averse to \textbf{non-English} ($\downarrow 78\%$). This aligns with the idea that internal monologue supports intuitive, self-directed thinking, making concise annotations preferable for minimizing cognitive load. Conclusion accuracy was also significantly higher for \lav{low volume} \textbf{English} variants ($\uparrow 28\%$) but lower in \textbf{non-English} ($\downarrow 32\%$, $\overline{\beta} = 0.74$, $\overline{f^2} = 0.55$). \full{Full-text} rankings dropped ($\downarrow 19\%$) in \textbf{English} but showed no significant change in \textbf{non-English}. \hav{High annotation volumes} increased across all languages, but the ranking changes were not found to be significant.

We also noted that when participants balanced their internal monologue across both languages, annotation preferences were more evenly distributed $\lav{\blacksquare} \hav{\blacksquare} \full{\blacksquare}$, reducing ranking variance ($\sigma^2 = 0.16$ vs. $\sigma^2 = 0.42$ in single-language thinkers).

\textbf{More frequent language mixing weakly decreases the preference for English annotations.} Language mixing is also known as code-switching/language alternation, and refers to the practice of using elements from two or more languages within a single conversation or utterance~\cite{deuchar2020code}. It is commonly seen in bilingual or multilingual individuals. Higher code-switching led to a non-significant decrease in English annotation preference ($\downarrow 8\%$) and conclusion selection accuracy ($\downarrow 14\%$), while increasing non-English preference ($\uparrow 9\%$) and conclusion selection accuracy ($\uparrow 13\%$), though effects were small $\lav{\blacksquare} \hav{\blacksquare} \full{\blacksquare}$ ($\overline{\beta} = -0.19$, $p = 0.07$; $\overline{\beta} = 0.13$, $p = 0.06$).

\textit{Hence, these results partially support \textbf{H3b}, as internal monologue balances preferences when conducted to an equivalent extent in both English and non-English, while language mixing frequency does not have any significant effects.
}

\textbf{Longer exposure to English-medium education slightly increases preference for English annotations.} 
English-medium education duration had a weak, non-significant effect on \textbf{English} annotation preference ($\uparrow 10\%$ rankings, $\uparrow 12\%$ accuracy, $\overline{\beta} = 0.27$, $\overline{f^2} = 0.14$), while reducing preference for \textbf{non-English} ($\downarrow 14\%$ rankings, $\downarrow 10\%$ accuracy). This trend was consistent across all chart types, with \textbf{scatterplots} and \textbf{heatmaps} showing a slightly stronger shift toward \textbf{English} annotations ($\uparrow 17\%$ rankings).

\textit{Hence, \textbf{H3c} is not supported, as longer durations of English-medium education do not significantly increase the preference for English annotations in bilingual participants.}

\vspace{-3mm}

\section{Qualitative Results}
\label{sec:7}

\begin{table*}[!ht]
\centering
\scriptsize
\resizebox{\textwidth}{!}{%
\begin{tabular}{@{}lll@{}}
\toprule
\textbf{Dimension}           & \textbf{Top Mentioned Codes (Count)}                                                                                                                                                                                                                                                 & \textbf{Examples of short phrases tagged}                                                                                                                                                                                                                                           \\ \midrule
\textbf{Trust}               & \begin{tabular}[c]{@{}l@{}}reliable, trustworthy, clear, credible, believable, dependable, familiar, \\ consistent, confident, used to, clear, context, helps, reinforce, relevant, \\ accurate, authentic, transparent, honest, valid, genuine, authoritative, certain\end{tabular} & \begin{tabular}[c]{@{}l@{}}"I believe the information presented."\\ "I feel confident in this data."\\ "This visualization seems very reliable."\\ "I trust this because it’s well explained."\\ "This looks like it’s coming from a trustworthy source."\end{tabular}              \\
\textbf{Engagement}          & \begin{tabular}[c]{@{}l@{}}interesting, engaging, attention-grabbing, thought-provoking, intuitive, \\ think about,  captivating, stimulating, holding, fascinating, enjoyable, \\ motivating, involving, compelling\end{tabular}                                                    & \begin{tabular}[c]{@{}l@{}}"I found myself thinking about it more deeply."\\ "It sparked my curiosity."\\ "I felt drawn into the data."\\ "This really made me reflect on the data."\\ "I could spend a long time studying this."\end{tabular}                                      \\
\textbf{Design Quality}      & \begin{tabular}[c]{@{}l@{}}easy to use, well-designed, clean, organized, aesthetically pleasing, \\ functional, professional, nice, likable,  neat, intuitive, user-friendly, \\ visually appealing, streamlined, well-crafted, high-quality, clear, \\ elegant, modern\end{tabular} & \begin{tabular}[c]{@{}l@{}}"The layout is clean and easy to navigate."\\ "It’s visually concise and clear."\\ "The design feels polished and professional."\\ "It’s well-designed and makes sense."\end{tabular}                                                                    \\
\textbf{Cognitive Load}      & \begin{tabular}[c]{@{}l@{}}overwhelming, complicated, simple, hard to process, heavy, easy to follow,  \\ simple, intuitive, challenging, burdensome, confusing, taxing, clear, \\ effortless, manageable\end{tabular}                                                               & \begin{tabular}[c]{@{}l@{}}"I had to spend extra time to understand."\\ "It was hard to keep track of all the elements."\\ "It was mentally exhausting."\\ "I felt confused by the layout and content."\\ "It’s straightforward and easy to understand."\end{tabular}               \\
\textbf{Language Complexity} & \begin{tabular}[c]{@{}l@{}}complicated, technical, easy, fluent, difficult to understand, clear, complex, \\ accessible, clear, concise, straightforward, convoluted, wordy, fluent, \\ dense, jargon-heavy\end{tabular}                                                             & \begin{tabular}[c]{@{}l@{}}"It was hard to follow because of the technical terms."\\ "The explanation is very straightforward."\\ "I had to re-read parts to fully understand."\\ "The annotations are too dense and long."\\ "The terms used are not familiar to me."\end{tabular} \\
\textbf{Objectivity}         & \begin{tabular}[c]{@{}l@{}}objective, unbiased, neutral, fair, balanced, accurate, impartial, detached, \\ accurate, factual, dispassionate, reasoned.\end{tabular}                                                                                                                  & \begin{tabular}[c]{@{}l@{}}"The data looks presented without any slant."\\ "It’s clear that this is impartial."\\ "I feel it’s represented fairly and accurately."\\ "It seems to be purely factual and detached."\\ "I appreciate how balanced the presentation is."\end{tabular}  \\ \bottomrule
\end{tabular}%
}
\caption{Examples of the most frequent words and phrases encountered for different dimensions of qualitative coding.}
\label{tab:coding}
\vspace{-5mm}
\end{table*}

Three annotators who participated in the stimuli design process qualitatively coded participants' free responses on their ``likes and dislikes" about the stimuli to explore the underlying cognitive and emotional dimensions behind language preference and performance patterns observed in the study. The responses were independently reviewed by two coders to ensure reliability. The initial coding involved marking keywords and recurring phrases. Afterward, the coders met to resolve discrepancies and refine the coding categories, ensuring that each response was appropriately categorized. The intercoder reliability, calculated using Cohen’s Kappa, yielded an agreement rate of 0.86, indicating strong reliability in the coding process. The coding was structured around three primary dimensions: \textit{trust}, \textit{engagement}, and \textit{design-quality} judgments. \anj{These dimensions were selected to provide a structured way of interpreting participants' perceptions and to explore how their reactions aligned with the quantitative results, in line with prior work done by Arunkumar et al. and Pandey et al.~\cite{arunkumar2023image,10669803,pandey2023you}.} Beyond these three categories, the analysis also included codes related to cognitive load, language complexity, and perceived objectivity to capture nuances in participants’ interactions with different annotations. Table~\ref{tab:coding} represents some of the top codes (mentioned more than 100 times by study participants), pertaining to each major dimension. The following sections present the key findings from this analysis.

\textbf{Trust in English annotations stems from clarity, while non-English text overloads.} Participants consistently found that detailed annotations in English, particularly at higher semantic levels (L3/L4), fostered greater trust due to increased clarity. One participant (English: title+3A) shared, ``I felt more confident about my understanding of the chart as there was more context. It just seemed clearer.'' This sense of trust was linked to the comprehensive nature of English annotations, where 204 participants (19\%) specifically cited context and relevance as key reasons for their increased trust in the visualizations.

However, when similar levels of detail were presented in non-English text, participants expressed a sense of overload. A total of 258 participants (23.5\%) mentioned that detailed annotations in non-English text were distracting, and 57 (5.2\%) specifically noted difficulty integrating L3/L4 content into their interpretation of the chart. One participant explained (Tamil: title+2A), ``Long sentences for background information are tiring to read and unnecessary and even distracting.'' In these cases, participants preferred shorter annotations (L1/L2), which were perceived as more neutral (noted by 122 participants, 11.1\%) and objective (73 participants, 6.6\%). Another participant remarked (Arabic: title+1A), ``This text didn’t try to explain too much, which made the chart feel more neutral and trustworthy.''

We posit that these trust dynamics stem from participants' exposure to English as a language of technical and academic discourse, where detailed explanations are common, versus the cognitive load introduced by long, complex sentences in non-English text.

\textbf{Simpler non-English annotations improve engagement by reducing perceived effort.} In contrast to the trust-related benefits of detailed annotations in English, engagement was higher when non-English annotations were simpler and less text-heavy. A total of 432 participants (39.4\%) noted that shorter annotations in their native language helped them stay focused on the data. ``These short phrases point me to focus more on data,'' one participant explained (Arabic: title+1A). Additionally, 382 participants (34.8\%) reported that higher annotation volumes in non-English text made the charts feel cluttered, reducing engagement.

Participants described how the complex grammatical structures in non-English text, such as Tamil's subject-object-verb order and Arabic's requirement to conjugate verbs for gender and number, increased perceived effort and reduced their ability to engage with the data. One participant shared (Tamil: title+2A), ``There need to be so many long extra words to explain how something looks that I can see already,'' while another mentioned (Tamil: title+3A), ``How do I quickly understand the data because of how long the sentences are?''

Conversely, in English annotations, participants were more engaged with longer annotations that included higher-level semantic content (L3/L4). One participant noted (English: title+2A), ``More explanation for these complex grids is good and walks me through everything,'' reflecting the higher engagement ratings observed in the quantitative analysis for detailed English annotations.

However, for non-English charts, higher annotation volumes had the opposite effect. In 205 instances (18.7\%), participants described non-English charts with large amounts of text as confusing, and 246 participants (22.4\%) found them overly complex. One participant said (Tamil, title+3A), ``The chart is trying to say too much in a small space,'' while another shared (Arabic, title+2A), ``The Arabic felt cluttered and hard to read.''

This effect was particularly pronounced for scatterplots and heatmaps, where 628 participants (57.2\%) reported increased reliance on text to interpret the data, especially in non-English annotations. As a result, the perceived cognitive effort was likely higher, leading to more frequent reports of confusion and disengagement. ``For scatter, I needed to read all the text just to understand the trend,'' one participant explained (Tamil, title+2A). This aligns with the quantitative findings, where the most significant effect sizes were observed for these chart types, indicating a greater reliance on text.

Additionally, 92 Arabic-speaking participants (17.1\%) reported that the left-to-right flow of the visualizations interfered with their ability to engage with the annotations. Arabic is typically read from right to left, and participants struggled with the mismatch between the flow of the text and the design of the charts. One participant noted (Arabic, title+1A), ``I kept wanting to start reading from the right, but the text and the dots went the other way, which made it harder to follow.'' This mismatch likely contributed to the perception that non-English charts were less well-organized, leading to lower design-quality ratings for these visualizations.

This difference in design perception between English and non-English text may stem from how languages are structured. For example, in Arabic, adjectives follow the noun they describe, making sentences feel longer and more difficult to process. One participant explained (Arabic, title+1A), ``The text makes the chart feel less organized and harder to follow.''
Thus, while higher annotation volumes in English enhanced participants' sense of design quality, they detracted from it in non-English charts by overloading readers with complex linguistic structures.

\textbf{Conclusion accuracy hampered by complex non-English annotations, especially in high-volume text.} Participants reported difficulties in accurately selecting conclusions when annotations in non-English text were complex or did not directly align with the data on the chart. In total, 98 participants in the Tamil (17.6\%) condition and 116 participants in the Arabic condition (20.1\%) struggled to validate conclusions that were not explicitly reflected in the annotations. One participant explained (Arabic, title+2A), ``It was harder to match when the text didn’t directly describe the data.''

Additionally, higher annotation volumes in non-English text exacerbated these difficulties. A total of 154 participants (14.1\%) found it challenging to identify correct conclusions in non-English charts with high annotation volumes, compared to only 34 participants (3.1\%) who experienced similar difficulties with English full-text annotations. One participant (Tamil, title+2A) commented, ``I found it harder to pick the right choice because there was just too much going on.''
This reinforces the quantitative results, where non-English annotations with high volumes of text led to lower accuracy in conclusion tasks. The increased cognitive load associated with processing complex non-English annotations likely impaired participants' ability to correctly interpret the key insights from the visualizations.

\section{General Discussion}
\label{sec:8}

\anj{This study examined how bilingual individuals interact with and prefer annotated visualizations across multiple languages and chart types. Our findings strongly validate the framework proposed by Stokes et al.~\cite{stokes2022striking}, demonstrating that annotation density and semantic content significantly influence comprehension accuracy. Consistent with their results, we observed that higher annotation density and rich semantic content (L3, L4) improved comprehension, particularly in charts with high information density.}

\anj{Extending this work, our study introduces a multilingual perspective, revealing that annotation preferences and comprehension patterns vary with language, summarized in Table~\ref{ressumm}. We provide the first evidence that when presented with charts in their native languages, participants preferred minimal annotations or full-text narratives, while English annotations were preferred for their succinctness in complex charts. Interestingly, native-language full-text narratives led to the highest comprehension accuracy, whereas fragmented annotations containing L4-level content introduced cognitive challenges. These results emphasize the need to balance annotation design, language processing, and cognitive load to optimize multilingual visualization strategies.}

\anj{In the following sections, we discuss the interplay between chart structure, annotation design, and language, highlighting their influence on preferences, engagement, and performance. We also explore implications for future multilingual visualization design.}

\begin{table}[!ht]
\centering
\scriptsize
\begin{tabular}{@{}p{0.1\linewidth}p{0.35\linewidth}p{0.1\linewidth}p{0.35\linewidth}@{}}
\toprule
\textbf{Research Question}    & \textbf{Finding}                                                                                         & \textbf{Languages}       & \textbf{Implications}                                                                                           \\ \midrule
\textbf{RQ1: Preferences}     & \textbf{English annotations}: High-density annotations (e.g., Title + 3A) were preferred, particularly for complex charts (e.g., scatterplots, heatmaps). & English             & Familiarity with English in technical contexts drives preferences for detailed, high-context annotations.       \\
                              & \textbf{Tamil/Arabic annotations}: Full-text annotations were favored, while fragmented annotations (e.g., Title + 3A) ranked lower. & Tamil, Arabic       & Narrative formats in native languages aid processing; fragmented annotations in native languages increase cognitive effort. \\
                              & Minimal annotations (e.g., Title-only) were moderately preferred for less information-dense charts (e.g., pie charts, bar charts). & Tamil, Arabic       & Simpler annotations align with reduced cognitive load for native-language visualizations.                      \\ \midrule
\textbf{RQ2: Comprehension}   & Increasing annotation density improved comprehension, with full-text paragraphs yielding the highest accuracy in all languages. & All Languages       & Full-text annotations facilitate linear processing, particularly in native languages for complex data.          \\
                              & Detailed annotations (e.g., Title + 3A) improved comprehension in English but showed mixed effects in Tamil/Arabic for high-density charts. & English, Tamil, Arabic & Semantic complexity (L3/L4) in native annotations can distract from data-centric interpretation, reducing accuracy. \\
                              & Text reliance increased with chart complexity (e.g., scatterplots, heatmaps) but did not moderate comprehension accuracy. & All Languages       & Complex charts benefit from robust annotations, though reliance on text does not guarantee comprehension gains. \\ \midrule
\textbf{RQ3: Individual Differences} & Fluency in English and exposure to English media correlated with a stronger preference for English annotations. & English             & Visual designs should consider the dominant language of users in academic or technical contexts to enhance engagement. \\
                              & Higher native-language immersion (e.g., frequent code-switching) balanced preferences across languages. & Tamil, Arabic       & Multilingual audiences benefit from flexible designs that accommodate both native and second-language fluency. \\ \bottomrule
\end{tabular}
\caption{\anj{Summary of findings from result analysis.}}
\label{ressumm}
\end{table}

\vspace{-5mm}

\subsection{Chart Reading vs. Concept Consolidation}

Our findings reveal distinct patterns in how bilingual participants engaged with visualizations annotated in English versus their native languages, suggesting that preferences are shaped by broader factors like language immersion, educational exposure, and familiarity with technical contexts. Bilingual participants preferred English annotations, likely due to greater academic and technical exposure in English (Table~\ref{demographics}). Those who used English for internal monologue also showed stronger preferences, aligning with research on language immersion and cognitive flexibility~\cite{bialystok2009bilingualism,deuchar2020code}.

In contrast, participants reading Tamil or Arabic annotations relied more on visual elements rather than textual summaries (L2), likely reflecting differences in educational exposure to data visualizations in their native language as well as prevalent informal native language usage contexts. While language immersion correlates with annotation preference, our findings do not provide causal evidence for why certain languages might influence concept consolidation differently; this may stem from familiarity with English as a technical language rather than inherent language differences. Overall, bilingual participants use their dominant language for familiar contexts while leveraging their native language for complementary understanding in visualization interpretation.

\subsection{Chart Structure Shapes Annotation Preferences}

Annotation preferences varied by chart type, reflecting the interplay between familiarity, annotation design, and cognitive load. Scatterplots and heatmaps imposed higher cognitive demands~\cite{cleveland1987graphical,quadri2024you}, leading to a strong preference for English annotations; given that all participants reported comparable familiarity with the chart types used in the study, this preference for English annotations is likely due to the succinctness of English text in managing high information density, rather than differences in chart familiarity.

The trade-off between succinctness and completeness, therefore, emerges as a critical factor in annotation design. Tamil and Arabic annotations were longer on average compared to their English counterparts due to the translation process, which expanded the text to maintain semantic equivalence. This in turn increased visual density, particularly in high-complexity charts like scatterplots and heatmaps, where space is limited~\cite{cleveland1987graphical,heer2007voyagers,tufte1985visual}. Higher visual density can raise cognitive load, making shorter annotations more effective for these charts. In contrast, bar and pie charts had lower visual density, allowing longer Tamil/Arabic annotations to fit without interfering with visuals~\cite{ware2019information,tufte1991envisioning}.

Additionally, Arabic speakers faced alignment challenges in axis-based charts (line, bar, scatterplot, heatmap), which may have influenced reduced preference for Arabic annotations~\cite{yi2021application}. This interference extended to axis-free charts (maps, pie charts), suggesting that layout design broadly influences annotation preferences. These findings underscore the importance of designing annotations that balance succinctness and completeness while considering the cognitive demands of chart types, the spatial constraints of the visualization, and the linguistic and cultural contexts of their users.

\subsection{The Disconnect Between Preference, Engagement, and Performance}

Participants’ preferences, engagement, and performance often diverged when interacting with visualizations. English annotations were generally preferred, especially for complex charts like scatterplots and heatmaps, but the preference gap narrowed when non-English text was presented in lower volumes. This suggests that participants’ familiarity with English in academic and technical contexts supports processing dense, abstract content.

An intriguing finding was the decrease in rankings for L4-level annotations in native languages, mirroring results from Lundgard et al.\cite{lundgard2021accessible}, where blind and visually impaired (BLV) participants disfavored high-level semantic content due to increased cognitive effort\cite{plass2003cognitive,genc2013effect}. In our study, participants prioritized fundamental comprehension (L1/L2) over abstract insights, focusing on text length and linguistic familiarity with technical terms in their native languages. Further, qualitative analysis showed higher engagement with simpler non-English annotations, while performance (conclusion selection accuracy) remained higher with English annotations, particularly for dense charts requiring abstract reasoning, highlighting a complex relationship between preference, engagement, and performance. While participants were more engaged with concise non-English text, English was preferred for clarity and participants performed better with English in high-cognitive-load tasks. Future studies could therefore explore whether reducing cognitive demands at lower semantic levels enables better engagement with L3/L4 content in native languages.

\subsection{Language in Flux: The Role of Mixing and Education}

We expected language mixing and English-medium education to strongly influence annotation preferences (\textbf{RQ3}), based on research linking code-switching to cognitive flexibility\cite{deuchar2020code,bialystok2009bilingualism} and English-medium education to higher English proficiency in academic contexts\cite{bylund2012effects,bialystok2017bilingual,rakotondravony2023beyond}. However, these factors showed weaker-than-expected effects, suggesting a more nuanced relationship.

\textbf{Language Mixing}: Participants who balanced internal monologue between English and their native language showed more even annotation preferences (Sec.\ref{sec63}), consistent with findings that bilinguals adapt language use based on task demands\cite{bialystok2009bilingualism}. This flexibility may explain why language mixing was not a strong moderator-- bilingual participants switched between languages naturally rather than favoring one annotation language exclusively.

\textbf{English Education}: Even one year of English-medium education increased English annotation preference ($\uparrow 7\%$) and conclusion selection accuracy ($\uparrow 12\%$), but preferences for English and non-English annotations remained relatively balanced ($\sigma^2 = 0.84$). This suggests that while English proficiency supports technical tasks, participants' native-language reliance persists, depending on chart complexity and annotation level.

Thus, language adaptation based on task demands-- rather than intrinsic cognitive differences-- may drive annotation preferences. Future research could explore mixed-language annotations, such as using English for high-level summaries and native languages for detailed content, to optimize multilingual data visualizations.

\subsection{Limitations and generalization}

While our study offers valuable insights into bilingual visualization preferences, several limitations should be noted. First, the study primarily focused  on English-Tamil and English-Arabic bilinguals, limiting generalizability to languages with different syntactic and morphological structures. Future research should explore a broader linguistic range to assess whether visualization interpretation varies across language families. Additionally, linguistic immersion in our study reflects both subjective experiences (e.g., media consumption) and observable patterns, though these were self-reported. To mitigate variability, participants underwent training trials with comprehension checks in both English and their native language, ensuring they could effectively process visual and textual information before proceeding. Only those meeting the required performance threshold participated, reducing the influence of self-report biases. Future studies should incorporate objective fluency assessments alongside training trials to further enhance reliability.

Second, participants were college-aged STEM students, which may not represent wider age groups or varying technological literacy levels; cognitive flexibility and language processing may differ for older adults, younger children, or individuals with diverse educational backgrounds~\cite{10294209}. We also did not assess prior exposure to data visualization analysis in formal education, which may influence comprehension and annotation preferences~\cite{10294209}.

Additionally, our study focused specifically on textual annotations within visualizations, neglecting other important design elements such as color, layout, and interactivity. These elements could interact with language and culture to shape how viewers engage with and interpret visualizations, such as in the case of Arabic, which is a right-to-left (RTL) language.  Text alignment and layout challenges in RTL charts could impact readability and annotation preferences, particularly for axis-based visualizations (e.g., scatterplots, bar charts, heatmaps). Future studies should explicitly examine how RTL layouts interact with annotation clarity and usability, considering cognitive adaptation to bi-directional text processing. Investigating how these design considerations, in conjunction with language, affect comprehension and user experience would enhance the applicability of visualization design across different cultural contexts.

Another limitation relates to the specific topics of the charts used in the study. All visualizations covered neutral topics, limiting insights into how annotation preferences shift for complex, emotional, or culturally specific content. Additionally, our structured comprehension tasks prioritized text-driven accuracy, potentially biasing results toward text-heavy formats. Future research should incorporate open-ended tasks to assess exploratory insight generation and lateral thinking strategies beyond structured conclusion selection.

Finally, differences in word length, stroke density, and translation complexity may have influenced text readability and engagement. Future studies should better control for linguistic density to assess its role in viewer comprehension and preference.
\section{Recommendation and Conclusion}
\label{sec:9}

\anj{
This paper explores the impact of multilingualism on reader preferences for annotated data visualizations. Through a large-scale experiment with bilingual participants, we found that annotation volume and content preferences varied based on language and visualization type. Our results suggest that English annotations were often preferred for denser visualizations, particularly scatterplots and heatmaps. In contrast, native language annotations supported user engagement in less dense stimuli like bar and pie charts. We attribute this to participants’ varying familiarity with English in academic and technical contexts, annotation density, and the visual demands of different chart types. These findings emphasize the need for design strategies that consider linguistic diversity, cognitive load, and user engagement to create more inclusive and effective visualizations.}

\anj{To improve the accessibility of data visualizations for multilingual audiences, we propose the following design recommendations:
}

\anj{1. \textbf{Enable language switching within visualizations}: 
Based on \textbf{RQ1}, which explored bilingual preferences for annotations in native versus second languages, future visualizations should offer interactive options for users to toggle between languages for specific annotations. This flexibility would allow users to engage with content in their preferred language based on task complexity or familiarity. For instance, toggling might enable users to view summaries in English for high-level comprehension and switch to native languages for detailed insights. This strategy ensures inclusivity while accommodating diverse linguistic proficiencies and user preferences.}

\anj{2. \textbf{Design mixed-language annotations for diverse cognitive needs}: Findings from \textbf{RQ3} highlight the complex interplay of language mixing in visualization interpretation. Bilingual users often navigate between languages based on task demands-- favoring English for technical precision and their native language for contextual depth. To support cognitive flexibility, annotations could therefore adopt a mixed-language approach tailored to semantic content and task demands. For example, using English for concise summaries or labels (e.g., ``Population growth: 10\% increase") vs. native languages for explanatory details in tooltips or accompanying captions (e.g., ``This reflects migration trends influenced by urban employment opportunities").
Such designs could balance cognitive ease and comprehension, especially for bilingual users who navigate between languages for technical and contextual understanding.}

\anj{3. \textbf{Adapt annotation strategies to chart complexity and visual density}: \textbf{RQ2} also showed that increased annotation volumes improved comprehension in both English and native languages. Annotation design should hence address the cognitive load imposed by chart complexity, visual clutter, and annotation density. In dense visualizations like scatterplots and heatmaps, prioritize concise annotations and offload detailed explanations to interactive elements like tooltips or expandable text boxes. Visual cues such as arrows/icons can also replace verbose descriptions. Additionally, segmenting longer translated annotations across chart elements (e.g., distributing text along axes or clusters) can mitigate layout challenges and ensure readability, particularly in languages with inherently longer text structures.} 

\anj{4. \textbf{Include contextual summaries alongside visualizations}: Examining \textbf{RQ2}, participants showed higher comprehension for full-text paragraphs in their native language for complex charts; hence, for complex visualizations, providing contextual summaries in paragraph form below/alongside the chart is advisable. These summaries would complement in-chart annotations by offering narrative-style explanations of takeaways, reducing reliance on dense text within the chart itself. For multilingual audiences, summaries could be available in both English and native languages, allowing users to choose formats that best support their comprehension and engagement.}

\anj{There remain open questions about how cultural and linguistic diversity can be better integrated into visualization design. Our findings emphasize that annotation design should prioritize flexibility, user control, and strategies that address cognitive load and visual clutter. We encourage researchers and practitioners to explore and test these recommendations, advancing the development of more inclusive and effective multilingual visualization designs.}

\section{Appendices}
\label{sec:appendices_inst}

\section*{Supplemental Materials}
\label{sec:supplemental_materials}

All supplemental materials are available at \url{https://osf.io/ckdb4/}, released under a CC BY 4.0 license.
In particular, they include:
(1) Excel files containing the aggregate data for collected measures, (2) Full set of stimuli used in the study along with source data visualizations, (3) SEM Analysis results (factor loadings, chi-squared, regression coefficients, significance, Cohen's f-squared), (4) demographic data of participants from Experiments, and (5) a full version of this paper with all appendices.

\section*{Figure Credits}
\label{sec:figure_credits}

Figure 3a image credit-- U.S. Department of The Treasury, April 2012. \url{https://democrats-financialservices.house.gov/uploadedfiles/20120413_financialcrisisresponsechartstreasury.pdf}

\begin{acks}
The authors thank Arun Madapusi and M.P. Srinivasan for their help with conducting the experiments at Anna University and BITS Pilani- Dubai respectively. The  authors also thank Viji Arunkumar, Shreyansh Raj, Deebthik Ravi, Rakshita Madhavan, Shuchi Talati, Sriram Venkatesan for their valuable work in the translation and coding of stimuli.
\end{acks}

\bibliographystyle{ACM-Reference-Format}
\bibliography{00_sample-base}

\appendix

\end{document}